\documentstyle[prc,psfig,aps,twocolumn]{revtex}
\begin{document}
\newcommand{\be}{\begin{eqnarray}}
\newcommand{\ee}{\end{eqnarray}}
\newcommand{\e}{ {\rm e} }
\newcommand{\la}{\langle}
\newcommand{\ra}{\rangle}
\newcommand{\no}{\nonumber}
\newcommand{\sls}{\!\!\!/}
\newcommand{\abr}{\alpha\beta\rho}
\newcommand{\abrb}{\alpha\beta{\bar \rho}}
\newcommand{\abrp}{\alpha\beta;\rho}
\newcommand{\abRp}{\alpha\beta;R}
\newcommand{\abAp}{\alpha\beta;A}
\newcommand{\abA}{\alpha\beta A}
\newcommand{\abR}{\alpha\beta R}

\twocolumn[
\hsize\textwidth\columnwidth\hsize\csname @twocolumnfalse\endcsname

\draft
\title{Hard photon production from unsaturated quark gluon
       plasma at two loop level}

\author{D. Dutta, S. V. S. Sastry, A. K. Mohanty, K. Kumar and R. K. Choudhury}
\address{ Nuclear Physics Division,
Bhabha Atomic Research Centre,\\
Trombay, Mumbai 400 085, India}

\maketitle

\begin{abstract}

The  hard  photon  productions from bremsstrahlung and annihilation with
scattering that arise at two loop level are estimated from a  chemically
non-equilibrated quark gluon plasma using the frame work of Hard Thermal
Loop  (HTL)  resummed  effective thermal field theory. The rate of photon
production is suppressed due to unsaturated phase space compared to it's
equilibrium counterpart. However, the suppression is relatively  weaker
than  expected  from  the  kinetic theory due to an additional collinear
enhancement arising from the decrease in  thermal  quark  mass.  For  an
unsaturated  plasma, unlike  the effective one loop case, the reduction in
the effective two loop processes is found to be  independent  of  gluon
fugacity, but strongly depends on quark and anti-quark fugacities. It is
also  found  that,  since  the  phase  space  suppression is highest for
annihilation  with  scattering,  the  photon  production   is   entirely
dominated  by  bremsstrahlung  mechanism  at all energies. This is to be
contrasted with the case of the equilibrated plasma  where  annihilation
with  scattering  dominates the photon production particularly at higher
energies.
\end{abstract}


]
\narrowtext

\section {Introduction}
The  study  of  single  photon  production  at  relativistic  heavy  ion
collisions has gained momentum in recent years due  to  availability  of
experimental data from CERN, SPS and also the data expected shortly from
the   RHIC   experiments   at  BNL  \cite{wa80,wa98,dks1,sar,wang,dks2}.
Assuming the formation of a quark gluon plasma  (QGP),  the  theoretical
studies  utilize  the  effective field theoretical formulation with hard
thermal loop (HTL) resummation technique \cite{pis1,bra1,bra2,fre1,fre2}
to calculate the imaginary part of the photon self energy and hence the
photon rate. An  important
aspect  in  this  approach  is to distinguish hard momentum of order $T$
from soft momentum of order $gT$ where $g$ is the QCD coupling constant.
The propagation of soft momentum is connected with infrared  divergences
in  loops  and therefore the  propagators  need  to  be  dressed
to get finite result.
According to Braaten and Pisarski \cite{bra1,bra2}, for  soft  momentum,
instead  of  using  bare propagators and vertices, effective propagators
and vertices should be used.  This  method  has  been  adopted  for  the
calculation of the rate of soft dilepton production \cite{bra3} and hard
real  photon  production  due  to  annihilation
($q{\bar q} \rightarrow
\gamma g$) and QCD Compton ($qg \rightarrow q  \gamma  $,  ${\bar  q}  g
\rightarrow  {\bar q} \gamma$)
processes from a quark matter at one loop
level in the effective theory \cite{kap,bair1}. In this method, a cutoff
parameter $(gT \le k_c \le T)$ is introduced to distinguish between  the
soft  and  the  hard  quark  momentum  circulating  in  the  loop.
For  hard  real  photon  production,  it is sufficient  to  use  an  effective
propagator  (summed  over successive one loop insertions) for one of the
quark loops carrying momentum below the cutoff while the other  loops  and
vertices  can  remain  undressed. Above the cutoff, bare propagators and
vertices can be used and a loop correction must be inserted on the  hard
propagator.  When adding the soft and the hard contributions, the cutoff
dependence cancels out.

Alternatively, the rate of photon production  can  also  be estimated on
the basis of  relativistic  kinetic  theory  where  the  integration  is
carried  out  over  a phase space volume multiplied by the square of the
reaction  amplitude  and  the  appropriate  distribution  functions  for
initial  and final states. In this approach also a cutoff
$k_c$ for  integration
 can  be  introduced  so  that  the  soft  part  that  involves
divergence can be treated separately. It is interesting to note that the
total  photon  production  rate  can also be estimated directly from the
hard part using a lower cutoff parameter for integration equal to  twice
the  thermal  quark  mass $(k_c^2=2m_q^2)$ \cite{kap}. This approach has
been extended to calculate  photon  production  from    non-equilibrium
plasma  \cite{trax,dutta1} with the use of thermal quark mass and parton
distribution  functions  appropriate  for  a  non-equilibrium  situation
\cite{bair2}.   In   case   of  a  non-equilibrated  plasma,  additional
contribution    is    expected    from     the     pinch     singularity
\cite{bair2,alt1,bel1}.  However, it is shown in \cite{bair2} that pinch
contribution at soft momentum scale is subleading with  respect  to  the
dominant   HTL  contribution.  Similarly,  for  the  hard  scale,  pinch
singularity is absent due to restricted kinematics.

It may be mentioned here that in the above cutoff method, a bare
gluon propagator has been used even if the cutoff does not constrain the gluon to
be hard. Allowing the gluon to be soft, leads
to new physical processes that may contribute to the hard photon production.
Recently, it is shown by Aurenche {\it et al.} \cite{arun1,arun2} that significant
contribution comes from the bremsstrahlung and
a new process called annihilation
with scattering (AWS) that arise at two loop level in the effective theory
due to space like soft gluon exchange. Their results
can also be written in a way that
separates the phase space from the amplitude of the process producing the
photon. The magnitude of the amplitude usually becomes less when the number of
loops increases. However, it is found that, within the effective theory,
the phase space contribution
at one loop level turns out to be smaller than the two loop  due to
kinematical constraints. Both effects compensate so that two loop diagrams
in the effective theory also contribute at the dominant level.

The work of Aurenche {\it et al.} \cite{arun1,arun2} assumes  the plasma
to be in
equilibrium at temperature $T$. However, the  rate of photon production may be affected
significantly if the phase space remains unsaturated.
The present work extends the formulation of Aurenche {\it et al.}
to the
non-equilibrium QGP.
We estimate the photon production from bremsstrahlung and AWS processes
for a chemically unsaturated quark gluon
plasma. We restrict to the region of Landau damping part ($L^2<0$ where $L$
is the gluon four momentum), whereas the region $L^2>0$ for  hard gluon
exchange has been included in the one loop calculations in the effective theory.
In a subsequent work \cite{arun3,arun4}, Aurenche
{\it et al.} have shown that even the higher order contribution can not be
ignored for real photon production indicating
that the thermal real photon production in QGP is a non-perturbative mechanism.
On the other hand, in such situation the
applicability of HTL resummation technique which is based on perturbative
approaches, may
become questionable \cite{ste}. However, the purpose of the present work is not
to go into the above aspect in detail. Here, we only focus on the bremsstrahlung and
the AWS photon production from a chemically unsaturated quark gluon plasma
which is of significance at RHIC and LHC energies
\cite{biro,biro2,lev,chak,shur1,geig,dks3,wong,dutta2}.

First we consider  the bremsstrahlung and
AWS photon productions from an equilibrated QGP \cite{priv}. We  draw
similar conclusions as that of previous work \cite{arun2} that, within the
framework of effective theory, the two loop contributions
particularly due to AWS, compete with one loop contributions at all
energies.
For a chemically non-equilibrated  plasma, since the phase
space is unsaturated, the photon productions both
at effective one and two loops level are suppressed as compared to the equilibrated case.
Since the thermal quark mass decreases with fugacities, the collinear
enhancement for the effective two loop processes also goes up. As a consequence, the
suppression at the effective two loop level has been found to be weaker than expected.
Interestingly, the above suppression is independent of the gluon fugacity
and depends only on unsaturated quark and anti-quark distribution functions.
Further, it is noticed that the suppression for the AWS process is the highest
and the photon production is entirely
dominated by the bremsstrahlung mechanism particularly when the plasma is
strongly unsaturated. This is contrary to the case of equilibrium situation
where AWS is the dominant mechanism of photon production at higher energies.
This result can be qualitatively understood on the basis of relativistic kinetic theory
where  the AWS process has more incoming fermion lines compared to the
bremsstrahlung process.
We may mention here that based on kinetic theory
argument and using the equilibrium results, an extension
to non-equilibrium QGP has been discussed by Mustafa {\it et al.} \cite{mus}.
However,
their results are not in agreement with the present findings which are derived
using the formalisms given in \cite{arun1,arun2}.

The paper is organized as follows. We begin with the description of an
unsaturated plasma with a brief review of the photon production both at one
and two loop level in the effective theory in section II. In section III, we evaluate the bremsstrahlung and the
AWS photon production from an unsaturated quark gluon plasma. We
show that the imaginary part of the self energy can  be written in a
form which separates the amplitude of the reaction from the phase space
so that the use of kinetic theory can be justified  for non-equilibrium plasma.
We discuss the results of photon production both at one and two loop
levels in the effective theory in section IV. Finally, the conclusion and summary are presented in
section V. For effective two loop calculation, we follow the
Retarded/Advanced (RA)
formalism \cite{arun5,arun6} where the
propagators remain same as zero temperature field theory while the vertices are
redefined which include nonequilibrium distribution functions. It is shown in the appendix that
the redefined vertex has the same form as that of equilibrium case except
that the distribution functions need to be defined appropriately to represent
a non-equilibrium phenomena.

\section {General formalism}

We consider  a  thermalized  plasma  of quarks and gluons expected to be
formed during the collisions of two heavy ions at relativistic energies.
However, at RHIC and LHC  energies,  several  perturbative-inspired  QCD
models  \cite{wang1,eskola}  predict  the  formation  of an unsaturated
plasma with high gluon content \cite{shur1}. Such a  plasma  will  attain
thermal equilibrium in a short time $t_0 \approx 0.3 - 0.7$ fm, but will
remain  far  from  chemical  equilibrium  \cite{biro}. Since the initial
plasma is gluon dominated, more quark and anti-quark pairs will be needed  in
order  to achieve chemical equilibration. The dynamical evolution of the
plasma undergoing chemical equilibration was studied initially  by  Biro
{\it et al.} \cite{biro} and subsequently by many others \cite{dutta1,dutta2}
by  solving  the  hydrodynamical  equations  along  with  a  set of rate
equations governing chemical equilibrations. In this  work,  we  do  not
consider  the  hydrodynamical  evolution of the plasma, rather calculate
only the static rate of  real  photon  production  from  an  unsaturated
plasma.  Further,  we also assume an ideal situation where the plasma is
baryon free.

A chemically non-equilibrated  but thermally equilibrated plasma can be
described  by the Juttner  distribution
function for quarks (anti-quarks) and gluons, given by,
\begin{equation}
n_q(p_0)=\left\{\begin{array}{ll}
n_q(|p_0|)~, & p_0>0\nonumber\\
1-n_q(|p_0|)~, & p_0<0
\end{array}\right.\nonumber
\end{equation}
\begin{equation}
n_g(p_0)=\left\{\begin{array}{ll}
n_g(|p_0|)~, & \mbox{$p_0>0$}\nonumber\\
-[1+n_g(|p_0|)]~, & \mbox{$p_0<0$}.
\end{array}\right.
\label{juttner}
\end{equation}
where
\begin{eqnarray}
n_i(|p_0|)&=&\frac{\lambda_i}{e^{|p_0|/T}\pm \lambda_i}\nonumber \\
&=&\frac{1}{e^{(|p_0|-\mu_i)/T}\pm 1}
\end{eqnarray}

The fugacity factor
$\lambda_i,~(i=q,{\bar  q},g)$  is related to the chemical potential $\mu_i$
as $\lambda_i=e^{\mu_i/T}$. The plus and minus signs are meant for the Fermi-Dirac
and Bose-Einstein distributions respectively. At equilibrium, chemical potential vanishes and
$\lambda_i \rightarrow 1$. The distribution functions can also be factorized
in an approximate way as $n_i=\lambda_i n^{eq}$ where $ n^{eq}$ is the
distribution function at equilibrium. In this representation, $\lambda_i$
gives a measure of deviation from the corresponding Fermi-Dirac or Bose-Einstein
distributions. In case
of a baryon rich plasma, the quark and the anti-quark
distribution functions can also be described by the same Juttner distribution
functions. However, the chemical potential will have an additional component to account
for the finite baryon density of the plasma (See Refs.
\cite{dutta1,dutta2} and the appendix A for detail ). In the present case, since
the plasma is baryon free, the fugacity and the chemical potential
both refer only to the unsaturation properties of the plasma.

The thermal photon production rate from such a plasma can be related to the retarded polarization
tensor of the photon using thermal field theory \cite{wel1}. For real photons,
this relation gives the number of photons emitted per unit time
and per unit volume of the plasma as
\begin{equation}
  2E{{dR}\over{d^3q}}=
 - {{2}\over{(2\pi)^3 }}\;n_{_B}(E)\,
  {\rm Im}\,\Pi^{^{RA}}{}_\mu{}^\mu(E,{\vec q})\; ,
  \label{realphot}
\end{equation}
where $E=q_0$ is the energy of the emitted photon and
$\Pi$ is the  retarded self-energy at finite T.
This  relation is valid only at first order in the QED coupling constant
$\alpha$ but is true for all orders  of  the  strong  coupling  constant
$\alpha_s$,  as  it  has  been assumed that the produced photons emerged
from the matter without further scattering.
It may be mentioned here that the emitted photon of energy $E$ is not the
part of the heat bath and therefore, has no distribution of any form.
Thus $n_{_B}(E)$ in Eq. (\ref{realphot}) is just a prefactor which happens to
coincide with the form of Bose-Einstein distribution only when
the plasma is in full equilibrium.
The  photon production  due to QCD Compton and annihilation processes
can be estimated by evaluating the photon self energy from the
diagrams shown in Figure 1.
The  imaginary  part  of  the self energy can be obtained by cutting
the above diagram using thermal cutting rule\cite{kobes1,kobes2,gelis}.
It  may be mentioned here that the rate of photon production
evaluated from imaginary part of photon self energy with some
finite  order of loop expansion is equivalent to the relativistic
kinetic theory \cite{kap}. The self energy calculated upto $N$ loop level
for $m$ particles $\rightarrow$ $n$ particles + $\gamma$,
is equivalent to the kinetic theory estimates from
all reactions consistent with  $m+n\le N+1$.
\begin{center}
\begin{figure}[!h]
\begin{minipage}{3.5cm}
\psfig{figure=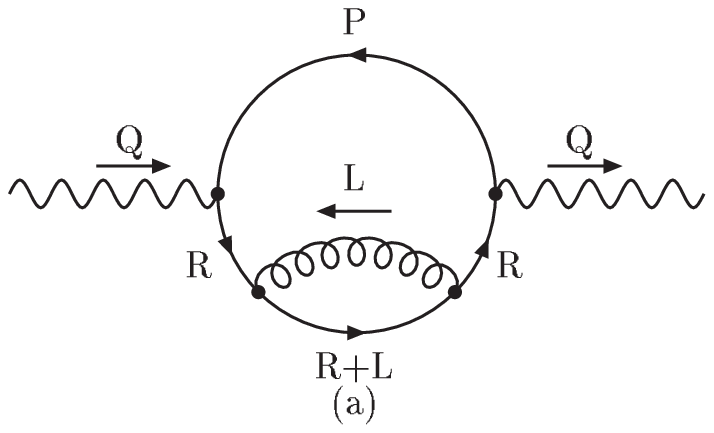,height=3.5cm,width=6cm}
\end{minipage}
\hspace{1cm}
\begin{minipage}{3.5cm}
\psfig{figure=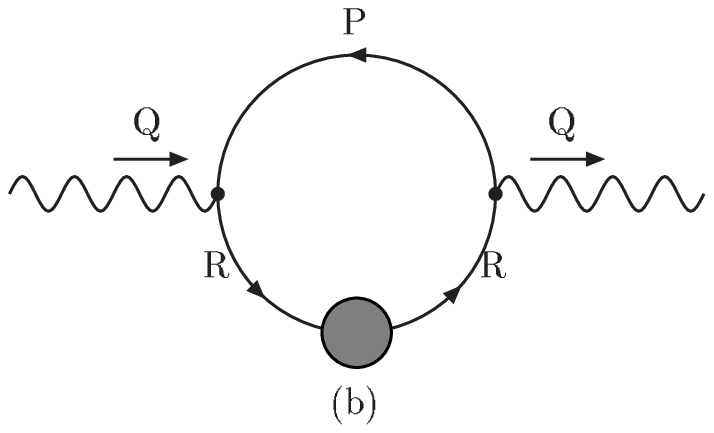,height=3.5cm,width=6cm}
\end{minipage}
\caption{Diagrams contributing to the photon self energy for
the Compton and annihilation processes when intermediate
quark momentum is (a) hard  (b) soft}
\label{One loop diagram}
\end{figure}
\end{center}

Similarly, the photon production from bremsstrahlung and
annihilation with scattering (AWS) processes
can be obtained from the
two loop effective diagrams as  shown in
Figure 2 (notations are same as given in \cite{arun1,arun2}).
\begin{center}
\begin{figure}[ht]
\begin{minipage}{3.5cm}
\psfig{figure=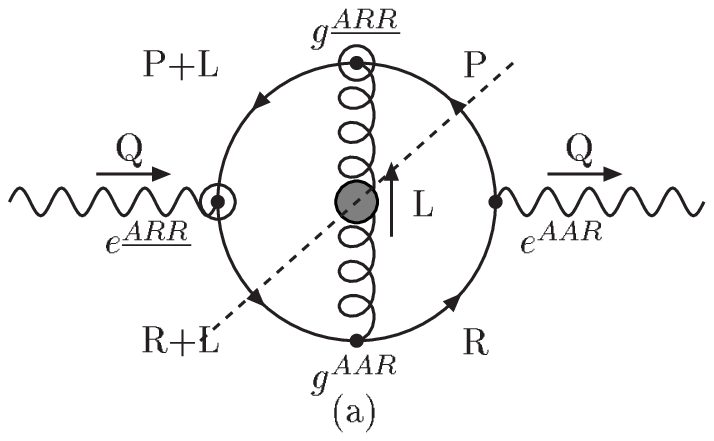,height=3.5cm,width=6cm}
\end{minipage}
\hspace{1cm}
\begin{minipage}{3.5cm}
\psfig{figure=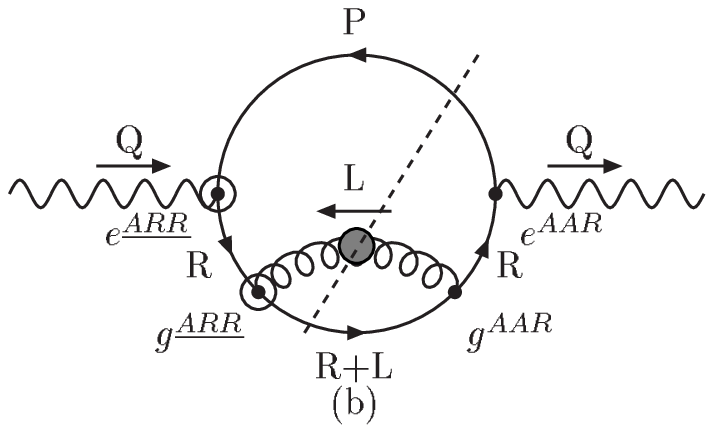,height=3.5cm,width=6cm}
\end{minipage}
\caption{(a) Vertex and (b) Self diagram for two loop self energy
in the effective theory}
\label{Two loop diagram}
\end{figure}
\end{center}
In section III, we estimate the above self energy (at effective two loop level)
more explicitly
for an unsaturated plasma as defined above.

\section {Photon production at two loop level}

In  the two loop level of effective theory the Feynmann diagrams need to
be considered are shown in Figure \ref{Two loop diagram}. The  imaginary
part  of  the  photon  self  energy  can  be expressed as a sum over the
possible cuts through the effective two  loop  diagrams.
The  physical  processes
bremsstrahlung  and quark anti-quark annihilation
are  obtained  by  cutting  through  the  effective   gluon
propagator.  To  obtain  the   contribution of bremsstrahlung and AWS,
the same order of
magnitude as of contribution from Compton and annihilation processes,
the quark momentum circulating in
the loop should be hard. As a result, all the vertices  and  propagators
can  be used as bare one except for the gluon propagator since the gluon
can be soft. Recall that only Landau Damping  part  ($L^2  <  0$)  gives
bremsstrahlung and AWS,   whereas   the   $L^2  >  0$  part  gives  Compton  and
annihilation  process  which  has been already  included  in the
effective one   loop   level
calculations for hard gluon exchange. However, due to phase space restriction,
the Landau damping is the dominant mechanism if $L$ is soft \cite{arun1}.

\begin{figure}[!h]
\centerline{\psfig{figure=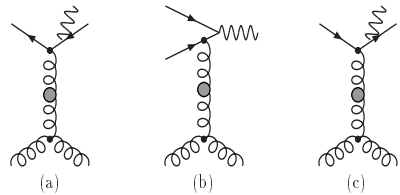,height=3cm,width=6cm}}
\caption{ Physical processes appear by cutting the two loop
diagram of Figure 2 for $L^2 < 0$ (a) Region I: (b) Region II (c) Region III}
\label{bremsstrahlung}
\end{figure}

Figure \ref{Two loop diagram} shows the relevant cuts and
circling required to evaluate the  self  energy  using  thermal  cutting
valid  for  RA  formalism  \cite{arun5,arun6,gelis}. In the following, we
extend the formalism of Aurenche {\it et al.} \cite{arun1,arun2}
to the chemically non-equilibrated situation.
An important aspect in the RA formalism is the redefined vertices which contain
the distribution functions. We have derived them in appendix A for a more
general situation where the plasma is chemically unsaturated as well as
has non zero baryo-chemical potential. However, in the present work, we consider
only a baryon free chemically unsaturated plasma. The vertex functions corresponding
to Figure 2(a) are given by
\begin{eqnarray}
&&e^{ AAR}(R,-P,-Q)=e[n_{_F}(r_0)-n_{_F}(p_0)]\nonumber\\
&&g^{ AAR}(R+L,-L,-R)=g[n_{_B}(l_0)+n_{_F}(r_0+l_0)]\nonumber\\
&&e^{\underline{ ARR}}(Q,P+L,-R-L)=-e^{ ARR}(Q,P+L,-R-L)\nonumber\\
&&~~~~~~~~~~~~~~~~~~~~~~~~~~~~~~~~~=-e\nonumber\\
&&g^{\underline{ ARR}}(-P-L,P,L)=-g^{ ARR}(-P-L,P,L)\nonumber\\
&&~~~~~~~~~~~~~~~~~~~~~~~~~~~~~~~~~=-g
\label{vertices}
\end{eqnarray}
The above vertex functions are in the same form as the equilibrated case except the distribution functions
should contain appropriate chemical potential. In the above example,
the chemical potentials associated with $R$ and $P$ lines are $\mu_q$ where as
it vanishes for $Q$ and $L$ lines. Therefore, using the above definitions,
the vertex (Figure \ref{Two loop diagram}(a)) and self
(Figure \ref{Two loop diagram}(b)) diagrams  of the imaginary part of the photon
self energy can be expressed as
\begin{eqnarray}&&
  {\rm Im}\,\Pi^{^{RA}}{}_\mu{}^\mu
  (E,{\vec q})_{|_{\rm vertex}}
  =-{\rm Im}\,\Pi^{^{AR}}{}_\mu{}^\mu
  (E,{\vec q})_{|_{\rm vertex}}\nonumber\\
  &&= {{NC_{_{F}}}\over 2}
  \int{{d^4P}\over{(2\pi)^4}}\int{{d^4L}\over{(2\pi)^4}}
  \nonumber\\
  &&\times\;
  e^{^{\underline{ARR}}}(Q,P+L,-R-L) g^{^{\underline{ARR}}}(-P-L,P,L)
  \nonumber\\
  &&\times\;
  g^{^{{AAR}}}(R+L,-L,-R)
  e^{^{{AAR}}}(R,-P,-Q)
  \nonumber\\
  &&\times\;
  {\rm Tr}\left[\gamma^\mu{\cal S}^{^{\underline{AR}}}(P+L)\gamma^\rho
    {\cal S}^{^{A\underline{R}}}(P)\gamma_\mu
  {\cal S}^{^{RA}}(R)\gamma^\sigma
    {\cal S}^{^{\underline{R}A}}(R+L)\right]\nonumber\\
  &&\times D^{^{A\underline{R}}}_{\rho\sigma}(L)
 \ee
 and
\begin{eqnarray}
  &&
  \!\!\!{\rm Im}\,\Pi^{^{RA}}{}_\mu{}^\mu
  (E,{\vec q})_{|_{\rm self}}=-
  {\rm Im}\,\Pi^{^{AR}}{}_\mu{}^\mu
  (E,{\vec q})_{|_{\rm self}}\nonumber\\
  &&= {{NC_{_{F}}}\over 2}
  \int{{d^4P}\over{(2\pi)^4}}\int{{d^4L}\over{(2\pi)^4}}
  \nonumber\\
  &&\times\;
  e^{^{\underline{ARR}}}(Q,P,-R)
  g^{^{\underline{ARR}}}(R,L,-R-L)
  \nonumber\\
  &&\times\;
  g^{^{{AAR}}}(R+L,-L,-R) e^{^{{AAR}}}(R,-P,-Q)
  \nonumber\\
  &&\times\;
  {\rm Tr}\,\left[\gamma^\mu{\cal S}^{^{A\underline{R}}}(P)\gamma_\mu
    {\cal S}^{^{{RA}}}(R)\gamma^\rho\right.
  \left.{\cal S}^{^{\underline{R}A}}(R+L)\gamma^\sigma
    {\cal S}^{^{\underline{RA}}}(R)\right]\nonumber\\
  &&\times D^{^{A\underline{R}}}_{\rho\sigma}(L)\; .
 \ee
According to the cutting rule valid for RA formalism \cite{gelis},
the fermion cut propagators are given by
\be
&&{\cal S}^{^{RA}}(P)\equiv {\cal S}^{^{A}}(P)\nonumber\\
&&{\cal S}^{^{\underline{AR}}}(P)\equiv -{\cal S}^{^A}(P)\nonumber\\
&&{\cal S}^{^{A\underline{R}}}(P)\equiv {\cal S}^{^R}(P)-{\cal S}^{^A}(P)\nonumber\\
&&{\cal S}^{^{\underline{R}A}}(P)\equiv {\cal S}^{^A}(P)-{\cal S}^{^R}(P).
\ee
We denote the fermion propagator:
\begin{eqnarray}
  &&{\cal S}^{^{R,A}}(P)\equiv{\overline{ P \sls}} S^{^{R,A}}(P){\ \rm with\ }
  {\overline{P}}\equiv(p_o,\sqrt{p^2+M^2_\infty}\,\hat{p})\\
  &&S^{^{R,A}}(P)
  \equiv{i\over{{\overline{P}}^2\pm ip_o\varepsilon}}
  ={i\over{P^2-M^2_{\infty}\pm ip_o\varepsilon}}\; .
  \end{eqnarray}
Similarly, boson cut propagator
\be
D_{\rho\sigma}^{^{A\underline{R}}}(L)\equiv D_{\rho\sigma}^{^R}(L)
-D_{\rho\sigma}^{^A}(L)\; .
\ee
 The effective gluon propagator in a linear covariant gauge is given by
 \begin{eqnarray}
  &&-D_{\rho\sigma}^{^{R,A}}(L)\equiv
  P^{^{T}}_{\rho\sigma}(L)\Delta^{^{R,A}}_{_{T}}(L)
  +P^{^{L}}_{\rho\sigma}(L)\Delta^{^{R,A}}_{_{L}}(L)\nonumber\\
  &&~~~~~~~~~~~~~~~~~~+\xi L_\rho L_\sigma/L^2\nonumber\\
  &&\Delta^{^{R,A}}_{_{T,L}}(L)\equiv
  \left.{i\over{L^2-\Pi_{_{T,L}}(L)}}\right|_{_{R,A}}\; .
 \ee
Here $P^{^{T,L}}_{\rho\sigma}$ are the usual transverse and longitudinal
projectors in linear covariant gauges \cite{arun2,wel2,land} and
  \be
  &&\Pi_{_{T}}(L)\equiv3m_{\rm g}^2\left[
    {{x^2}\over 2}+{{x(1-x^2)}\over{4}}\ln\left(
      {{x+1}\over{x-1}}\right)\right]\\
  &&\Pi_{_{L}}(L)\equiv3m_{\rm g}^2(1-x^2)\left[
    1-{{x}\over{2}}\ln\left(
      {{x+1}\over{x-1}}\right)\right]\; ,
\end{eqnarray}
where, $x=l_0/l$.

Using the expression of the vertices in Eq.(\ref{vertices}) and
the cut propagators, the vertex diagram
of photon self energy can be simplified as
  \be
  &&{\rm Im}\,\Pi^{^{RA}}{}_\mu{}^\mu
  (E,{\vec q})_{|_{\rm vertex}}\nonumber\\
  &&=-{{NC_{_{F}}}\over 2}e^2g^2
  \int{{d^4P}\over{(2\pi)^4}}               \int{{d^4L}\over{(2\pi)^4}}
  \left[\Delta^{^{R}}_{_{T,L}}(L)     -     \Delta^{^{A}}_{_{T,L}}(L)\right]
  \nonumber\\
  &&\times\;
  \left[S^{^{R}}(P)-S^{^{A}}(P)\right]
  \left[S^{^{R}}(R+L)-S^{^{A}}(R+L)\right]
  \nonumber\\
  &&\times\;
  \left(n_{_{F}}(r_o)-n_{_{F}}(p_o)\right)
  \left(n_{_{B}}(l_o)+n_{_{F}}(r_o+l_o)\right)
  \nonumber\\
  &&\times\;
  S(R)S(P+L)P_{\rho\sigma}^{^{T,L}}(L)\;
  {\rm Trace}^{\rho\sigma}{}_{|_{\rm vertex}}\; .
  \label{cutvertex}
\end{eqnarray}
Similarly, the photon self energy for the self diagram
can be expressed as in Ref. \cite{arun2}
with appropriate distribution functions
  \be
&&  \!\!\!{\rm Im}\,\Pi^{^{RA}}{}_\mu{}^\mu
  (E,{\vec q})_{|_{\rm self}}\nonumber\\
  &&=-{{NC_{_{F}}}\over 2}e^2g^2
  \int{{d^4P}\over{(2\pi)^4}}\int{{d^4L}\over{(2\pi)^4}}
  \left[\Delta^{^{R}}_{_{T,L}}(L)-\Delta^{^{A}}_{_{T,L}}(L)\right]
  \nonumber\\
  &&\times\;
  \left[S^{^{R}}(P)-S^{^{A}}(P)\right]
  \left[S^{^{R}}(R+L)-S^{^{A}}(R+L)\right]\nonumber\\
  &&\times\;
  \left(S(R)\right)^2
  \left(n_{_{F}}(r_o)-n_{_{F}}(p_o)\right)
  \left(n_{_{B}}(l_o)+n_{_{F}}(r_o+l_o)\right)\nonumber\\
  &&\times P_{\rho\sigma}^{^{T,L}}(L)\;
  {\rm Trace}^{\rho\sigma}{}_{|_{\rm self}}\;
  \label{cutself}
\end{eqnarray}
where
$e$ is the electric charge of the quark which depends on its flavor.
The trace in Eq. (\ref{cutvertex}) and Eq.(\ref{cutself}) are given by,
\be
{\rm Trace}^{\rho\sigma}{}_{|_{\rm vertex}}&=&{\rm Tr}[\gamma^\mu
{ ({P \sls} +{L \sls})}
\gamma^\rho({ P \sls})\gamma_\mu({ R \sls})
\gamma^\sigma({ R \sls}+{ L \sls})]\nonumber\\
{\rm Trace}^{\rho\sigma}{}_{|_{\rm self}}&=&{\rm Tr}[\gamma^\mu{ P \sls}
\gamma_\mu({ R \sls})\gamma^\rho({ R \sls}+ { L \sls})
\gamma^\sigma{ R \sls}]\; .
\ee
The factor $S(R)$, $S(P+L)$ without any $R$ and $A$
superscript denotes the principal part of the propagator i.e.,
\be
S(R)&=& \frac{1}{(R^2-M_\infty^2)}\nonumber \\
S(P+L)&=&\frac{1}{((P+L)^2-M_\infty^2)}\; .
\label{cutpropagator}
\ee
The difference between the retarded and the advanced gluon propagators,
 in  both the vertex (Eq. \ref{cutvertex}) and self (Eq. \ref{cutself})
 part of the photon self-energy is known as spectral function given by
 \be
 &&\rho_{_{T,L}}(L)=\Delta^{^{R}}_{_{T,L}}(L)-\Delta^{^{A}}_{_{T,L}}(L)\; .
 \ee
  The properties of these spectral functions depend upon the analytic
  structure of the gluon propagator. For $L^2<0$ ({\it i.e.} $|x|<1$) region,
  in which we are interested about, the self-energies $\Pi_{_{T,L}}$ acquire
  an imaginary part due to the logarithm and the corresponding expression
  of the spectral functions are given by,
  \be
  &&\rho_{_{T,L}}(L)\equiv \frac{\left.{-2{\rm Im}\Pi_{_{T,L}}(L)}\right|_{_{R}}}
  {(L^2-{\rm Re}\Pi_{_{T,L}}(L))^2 - (\left.{{\rm Im}\Pi_{_{T,L}}(L)}\right|_{_{R}})^2}\; .
  \ee
Here the imaginary part of retarded self energies are
\be
&&\left.{{\rm Im}\Pi_{_{T}}(L)}\right|_{_{R}}=\frac{3\pi m_g^2}{4}x(1-x^2)
\nonumber\\
&&\left.{{\rm Im}\Pi_{_{L}}(L)}\right|_{_{R}}=-2\left.{{\rm Im}\Pi_{_{T}}(L)}\right|_{_{R}}\; .
\ee
Adding the contributions from these two diagrams
and plugging into the expression (Eq.(\ref{realphot})),  the photon production
from the effective two loop level can be evaluated.
It should be noticed that the expression for the self energy is same as
used in \cite{arun2}, except for the non-equilibrium distribution
functions $n_{_F}$ and thermal masses which contain chemical potentials.
Like in one loop case, here also we ignore the pinch contribution and assume
that the effective quark and gluon propagators are still given by their
equilibrium counter part with use of asymptotic thermal quark mass
$M_\infty^2=2m_q^2$  defined by
\be
m_q^2=\frac{g^2}{3\pi^2}\int p~ dp~ [2n_g+(n_q+n_{\bar q})]
\label{thermal quark}
\ee
which can be shown to be $(\lambda_g+\frac{\lambda_q}{2})g^2T^2/9$ for
factorized distributions.
The thermal gluon mass $m_{\rm g}^2$ appropriate for non-equilibrium plasma
\cite{wel2,fle1} is given by
\be
m_{\rm g}^2=\frac{g^2}{3\pi^2}\int p dp [6n_g+N_f(n_q+n_{\bar q})]\; .
\ee
For a factorised distribution the $m_g^2$ can be written as
$(3\lambda_g+\frac{N_f}{2}\lambda_q)g^2T^2/9$.
The kinematic conditions restrict the phase space for the
physical process into three regions as shown in Figure \ref{bremsstrahlung}.
\[\begin{array}[!h]{cc}
\mbox{Region I} &p_0<0~~ \mbox{ and}~~ r_0+l_0<0  \\
\mbox{Region II} &p_0<0~~\mbox{ and}~~ r_0+l_0>0  \\
\mbox{Region III} &p_0>0~~\mbox{ and}~~ r_0+l_0>0 \\
\end{array}\]
Region I and III corresponds to bremsstrahlung from anti-quark
and quark respectively.
The region II corresponds to AWS processes.
The physical processes with
the  lower line in the Fig. \ref{bremsstrahlung} replaced with  quark,
anti-quark have also been included.
It may be mentioned here  that region I and region III will give
same contribution as long as   quark and anti-quark distribution
functions are same (i.e. for a baryon free plasma).
We will study the contribution
from region III and multiply it by a factor 2 to get the total
bremsstrahlung photon yield. Similarly, we will discuss about the
region II for AWS process.

\subsection{Bremsstrahlung }

The two points which need to be addressed here are the collinear limit and
the enhancement due to collinear singularity.
It is important to notice that due to the
factors $[(R^2-M_\infty^2)((P+L)^2-M_\infty^2)]^{-1}$
which appear in the integral Eq. (\ref{cutvertex})
is responsible for
the collinear divergence. Although we will carry out the complete
integration, even at the qualitative level, we can understand
the nature of the divergence from the integral of the type

\begin{equation}
I\equiv\int_0^2 \frac{du}{(R^2-M_\infty^2)} \int_0^{2\pi}
         \frac{d\phi}{(P+L)^2-M_\infty^2}
\end{equation}
where $\theta~(u=1-cos\theta)$ is the angle between
$\vec p$ and $\vec q$ and $\phi$ is the
azimuthal angle between  $\vec q$ and $\vec l$ when projected on a plane
orthogonal to $\vec r$. For hard photon production,
the following approximations can be used
\begin{eqnarray}
R^2-M_\infty^2 &\approx& 2pq(u+a);~~a=\frac{M_\infty^2}{2p^2}\nonumber\\
\int \frac{d\phi}{(P+L)^2-M_\infty^2} &\approx &\frac{2\pi (p+q)}{2qp^2(u+b)};~
b=\frac{M_\infty^2}{2p^2}+\frac{L^2}{2p^2}
\end{eqnarray}
In the above, $(b-a) \sim L^2/p^2$ gives a measure of the distance between
the two poles. Since the two poles are very near by (for soft gluon exchange),
the above integral is
of the order
\begin{eqnarray}
I \sim \frac{p+q}{p(p^2q^2)a}
\end{eqnarray}
for $a> L^2/p^2$ or $M_\infty^2>L^2$.
Note that this enhancement factor
which is associated with the smallness of the angle of emission
is same irrespective of whether the emitted photon is soft or hard.
As a consequence, the integral is enhanced by a
factor of order $p^2/M_\infty^2 \sim 1/(g^2\lambda_g)$ if the
plasma is strongly gluon dominated i.e. $\lambda_q<\lambda_g\ll 1$.
The above enhancement is
larger by a factor of $\lambda_g^{-1}$
as compared to the equilibrated case. Although the above argument is
qualitative, it
remains valid even after all the integrations are carried out
(see Figure 4 in section IV).
Now, proceeding as before \cite{arun2} under collinear
approximation, the expression for the imaginary part of self energy
can be written as
\begin{eqnarray}\lefteqn{
  {\rm Im}\,\Pi^{^{AR}}{}_\mu{}^\mu
  {}(E,{\vec q}) }\nonumber\\
  &&\approx(-1)_T\frac{NC_F}{4\pi^4}e^2g^2\frac{1}{E^2}\int_{p^\ast}^{\infty}
  [n_{_F}(p)-n_{_F}(p+E)]\nonumber\\
  &&\times(p^2+(p+E)^2)~dp
  \int_0^{l^\ast}l^4~dl\int_{-1}^{1}dx\int_0^1 du^\prime~ n_{_B}(lx)
  \nonumber\\
  &&\times\rho_{T,L}(l,lx)(1-x^2)^2(1-u^\prime )^{-1/2}
  (4M_\infty^2+l^2(1-x^2)u^\prime)^{-1}\nonumber\\
\label{self}
\end{eqnarray}
where $u^\prime\equiv -8r^2u/L^2$ and the symbol $(-1)_T$ denotes an
extra minus sign in the transverse contribution.
Here we have introduced some cut-offs $p^\ast$ and $l^*$ at a scale
intermediate between $gT$ and $T$, where we assume $r$ to be hard
and $l$ to be negligible compared to $T$.
We have  ignored the factor
$n_{_F}(r_0+l_0)$  since it is much smaller compared to the
Bose distribution $n_{_B}(l_0)$  (where $l_0=lx$) particularly for  unsaturated QGP.
As a result, the integral over $p$
becomes independent of integrals over $L$.
The  Bose-distribution is also approximated to
$n_{_B}(lx)\sim \frac{T}{ lx} $.
This assumption considerably simplifies the
numerical evaluation of the self energy which can be expressed
in terms of the dimensionless constants $J_T$ and $J_L$,
\begin{eqnarray}
J_{T,L}&=&\int_0^1 \frac{dx}{x}{\tilde I}_{T,L}(x) \nonumber \\
&&\times \int_0^{w^*} dw \frac{\sqrt{\frac{w}{w+4}}
\tanh^{-1}\sqrt{\frac{w}{w+4}}}
{(w+{\tilde R}_{T,L}(x))^2+({\tilde I}_{T,L}(x))^2 }
\label{jtjl}
\end{eqnarray}
where
\begin{eqnarray}
&&w\equiv \frac{-L^2}{M_\infty^2},~~~~~~~~
{\tilde I}_{T,L}(x)\equiv \frac{{\rm Im}\Pi_{T,L}(x)}{M_\infty^2},\nonumber\\
&&{\tilde R}_{T,L}(x)\equiv \frac{{\rm Re}\Pi_{T,L}(x)}{M_\infty^2}.\nonumber
\end{eqnarray}
Note that we do not assign any chemical
potential to the gluon line $L$ leading to the approximation
$\frac{T}{ lx} $ rather than $\lambda_g \frac{T}{ lx}$
(see subsection $D$ for more discussion).
The functions  $J_T,~ J_L$ depend on the thermal mass ratio
$m_{\rm g}^2/M_\infty^2$ and $l^\ast/M_\infty$.
Notice that the above expressions for $J_T$ and $J_L$ are same as in
Ref.\cite{arun2}
except that the thermal masses now depend on the chemical potentials or
fugacities.
It has been shown in \cite{arun1,arun2}
that for equilibrated plasma, taking $w^\ast\rightarrow \infty$
introduces a negligible contribution to the integration of $J_{T,L}$.
This argument is also valid for chemically non-equilibrated plasma
since the value of $w^\ast$ increases with decreasing fugacity.
Thus extrapolating the
upper limit of the integration $w^\ast$ to $\infty$ introduces
smaller error as compared to the equilibrated plasma and can be neglected.
Finally, the imaginary part of the self energy can be written as
\begin{eqnarray}\lefteqn{
  {\rm Im}\,\Pi^{^{AR}}{}_\mu{}^\mu
  {}(E,{\vec q}) }\nonumber\\
  &\approx &\frac{NC_F}{\pi^4}e^2g^2(J_T-J_L)\frac{T}{E^2}\nonumber\\
  &&\times\int_{0}^{\infty} [n_{_F}(p)-n_{_F}(p+E)](p^2+(p+E)^2)~dp
\end{eqnarray}
The total photon production rate for bremsstrahlung process
can  be evaluated as
\begin{eqnarray}
&&\left.2E\frac{dR}{d^3q}\right|_{brem}\approx 4 \frac{NC_F\alpha\alpha_s}{\pi^5}
\left(\sum_f e_f^2\right)~\frac{T}{E^2}(J_T-J_L)\nonumber\\
&&\times n_{_B}(E)~\int_0^\infty dp~(p^2+(p+E)^2) [n_{_F}(p)-n_{_F}(p+E)]
\label{ratejtjl}
\end{eqnarray}
where $e_f$ is the electric charge of the quark flavor $f$ in units
of electron charge and
$n_{_B}(E)=[\exp(E/T)-1]^{-1}$ is the prefactor independent of any chemical
potential for the bremsstrahlung process. The $J_T$ and $J_L$ integrals depend only on the
thermal mass ratio $m_g^2/M_\infty^2$ and are insensitive to the chemical potential or
fugacity. Since the prefactor $n_B(E)$ is independent of $\mu$ for bremsstrahlung,
the chemical unsaturation effect enters only through the
quark distribution functions $n_F(p)$ in the $p$
integral. Therefore, for factorized distribution functions, the bremsstrahlung
contribution from non-equilibrated plasma is suppressed by a factor of $\lambda_q$
as compared to the equilibrium case.

\subsection {Annihilation with scattering }

 In case of $q{\bar q}$ annihilation with scattering (AWS),
 one should consider the region II
 where $p_0<0$. The procedure to estimate AWS rate is nearly identical to
that of bremsstrahlung except for the exchanges $u \rightarrow v$
 and $p \rightarrow -p$
where $v= 1+\cos \theta$. In this case also the enhancement mechanism
remains same as before.
Finally, with the replacement of $p$ by $-p$,
 the $p$ integral in Eq.(~\ref{self}) can be written as
\begin{eqnarray}
&&\left.2E\frac{dR}{d^3q}\right|_{AWS}\approx 2 \frac{NC_F\alpha\alpha_s}{\pi^5}
\left(\sum_f e_f^2\right)~\frac{T}{E^2}(J_T-J_L)\nonumber\\
&&\times n_{_B}(E)~\int_0^\infty dp(p^2+(E-p)^2) [n_{_F}(-p)-n_{_F}(E-p)]
\label{ratejtj2}
\end{eqnarray}
where $n_{_F}(-p)=1-n_{_F}(p)$. For equilibrated plasma, the contribution from
$[n_{_F}(-p)-n_{_F}(E-p)]$ is assumed $\sim 1$. This approximation is also valid
in case of non-equilibrated plasma since the distribution functions are smaller
by a factor of $\lambda_q$. The unsaturation effect appears through
the prefactor $n_{_B}(E)$ which goes as $[\exp(E/T)-1]^{-1} \lambda_q^2$ under
factorized approximation
(see the discussions in appendix A).
Therefore, the AWS photon production is suppressed by a factor
of $\lambda_q^2$ as compared to its equilibrium counterpart.

Finally, we would like to end this sub-section with the remark that both bremsstrahlung
and AWS photon productions for non-equilibrated plasma are suppressed by
a factor of $\lambda_q$ and $\lambda_q^2$ respectively
due to unsaturated quark and
anti-quark distribution functions. The suppression due to unsaturated gluon
distribution function seems to get compensated by the additional
enhancement caused by collinear singularity.

\subsection {Factorisation and Kinetic theory}

In this section, we establish a connection between the field
theoretical formalism as given in previous sub-section and relativistic
kinetic theory which can also be used to estimate the photon production
rate under semiclassical approximation \cite{arun1,cley1}.
Consider the case of photon production due to bremsstrahlung.
In the kinetic theory approach,
the  photon production rate can be
evaluated by integrating the amplitude squared of the
process over the phase space of unobserved particles
and given by
\begin{eqnarray}
&&\frac{dR}{d^3q}=\frac{1}{(2\pi)^3~2E}\int\frac{d^4p}{(2\pi)^4}
\int\frac{d^4K}{(2\pi)^4} \int\frac{d^4L}{(2\pi)^4}\nonumber\\
&&\times\left|\raisebox{-.9cm}{\begin{picture}(130,15)(0,0){
\mbox{\psfig{figure=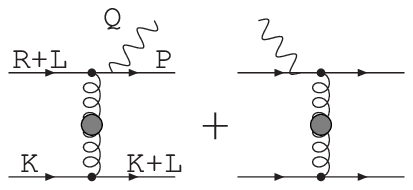}}
}\end{picture}}\right|^2\nonumber\\
&&\times 2\pi\delta(P^2-M_\infty^2)~2\pi\delta((R+L)^2-M_\infty^2)\nonumber\\
&&\times 2\pi\delta(K^2-M_\infty^2)~2\pi\delta((K+L)^2-M_\infty^2)\nonumber\\
&&\times n_{_F}(r_0+l_0)~n_{_F}(k_0)~[1-n_{_F}(p_0)]~[1-n_{_F}(k_0+l_0)]\nonumber\\
\label{rela1}
\end{eqnarray}
Here we have only considered the amplitude for the
bremsstrahlung process where quark has been scattered from
another quark. In order to get total photon production rate from
bremsstrahlung, the processes involving quark scattered
from a gluon or an anti-quark also have to be considered.

In order to establish a connection to field theory, we now
look for various statistical factors that appear in Eq.(\ref{rela1}). Let
us consider the product of all the distribution functions that appear
in the calculation of self energy [see Eq.(\ref{realphot}) and Eq.(\ref{cutvertex})]
\begin{eqnarray}
\lefteqn{n_{_B}(q_0)[n_{_B}(l_0)+n_{_F}(r_0+l_0)]}\nonumber\\
&&\times [n_{_F}(r_0)-n_{_F}(p_0)][n_{_F}(k_0+l_0)-n_{_F}(k_0)]\nonumber\\
&&= n_{_F}(r_0+l_0)[1-n_{_F}(p_0)] n_{_F}(k_0)[1-n_{_F}(k_0+l_0)]
\label{identity1}
\end{eqnarray}
The pre-factor $n_{_B}(q_0)$ ($q_0= E$) comes from Eq.(\ref{realphot}) while the second and third
factors
come from the vertex functions $g^{AAR}$ and $e^{AAR}$. Although not explicit
in Eq.(\ref{cutvertex}), the last factor appears due to the hard thermal quark loop contribution
to the gluon self energy \cite{arun1}. In the last factor, $n_{_F}$ will be replaced by $n_{_B}$
when the quark scattering from gluon is considered.
For the above identity to be valid, the distribution functions should be
defined properly with appropriate chemical potentials.
For example, in case of
bremsstrahlung, the chemical potential for $Q$ and $L$ lines are zero where as
chemical potentials for $R+L$, $R$ and $P$ lines are $\mu_q$. Similarly, the
chemical potentials for $K$ and $K+L$ lines are $\mu_q$ for quark loop and $\mu_g$
for gluon loop respectively (see appendix A for detail). The baryo-chemical potential is zero since we
restrict only to the case of a baryon free plasma. Due to the validity of the
above factorisation, similar to the  case of an  equilibrated
plasma \cite{arun1,arun2}, we can  show the equivalence between
two approaches based on
field theory and kinetic theory. Similar arguments
also apply for the case of annihilation with scattering due to the following
identity (given only for quark quark scattering)
\begin{eqnarray}
\lefteqn{n_{_B}(q_0)[n_{_B}(l_0)+n_{_F}(r_0+l_0)]}\nonumber\\
&&\times [n_{_F}(r_0)-n_{_F}(-p_0)][n_{_F}(k_0+l_0)-n_{_F}(k_0)]\nonumber\\
&&= n_{_F}(r_0+l_0)n_{_F}(p_0)n_{_B}(k_0)[1-n_{_F}(k_0+l_0)].
\label{identity2}
\end{eqnarray}
The above identity is quite similar to Eq.(\ref{identity1})
except that the
chemical potential associated with the pre-factor $n_B(q_0)$ is  $2\mu_q$.

Recently, Mustafa {\it et al.} \cite{mus} have calculated the photon production from a
non-equilibrated plasma based on the above kinetic theory approach.
Assuming a factorized form of parton distribution functions i.e. $n_i=\lambda_i n_i^{eq}$
where $ n_i^{eq}$ is the distribution function at equilibrium and
combining the contributions from quark and gluon scattering
with appropriate spin, colour and flavour statistics, the rate for the
bremsstrahlung production can be written as [see Eq.(\ref{rela1})].
\begin{eqnarray}
\frac{dR}{d^3q}&=&(\frac{A}{A+B}\lambda_g\lambda_q+
\frac{B}{A+B}\lambda_q^2)\times{\cal R}\nonumber\\
&=&\lambda_q(\frac{2}{5}\lambda_g+\frac{3}{5}\lambda_q)
\times{\cal R}
\label{factor}
\end{eqnarray}
where $\cal R$ is the equilibrium contribution to photon production,
${\cal R}=(A+B) \cal I$
with $\cal I$ given by Eq.({\ref{rela1}) to be evaluated under
classical approximation. The degeneracy
factors $A$ and $B$ are given by
\begin{eqnarray}
A&=&\frac{4}{9}\times 2\times 2_f\times 2_s\times 3_c
=\frac{32}{3}~~~;\nonumber\\
B&=&2_s\times 8_c = 16.\nonumber
\end{eqnarray}
The factor $4/9$ in the expression for $A$ appears due to the assumption
$|{\cal M}|^2_{q\leftrightarrow q}=4/9|{\cal M}|^2_{q\leftrightarrow g}$ where
$|{\cal M}|^2_{q\leftrightarrow q}$ and
$|{\cal M}|^2_{q\leftrightarrow g}$ are the square
of the matrix element for quark-quark
scattering and quark-gluon scattering respectively \cite{cley1}.
Recall that it is possible to combine the quark and gluon contribution in a
form given by Eq.(\ref {factor}) only if $n_{_F} \approx n_{_B}$ and quantum statistics are
ignored in Eq.(\ref{rela1}) which is true under classical approximation.
In this context, the factorisation in Eq.(\ref{factor}) is only approximate.
This has been the basis of the result used in Ref. \cite{mus}, although
they  subsequently use correct equilibrated value for $\cal R$ obtained from
the imaginary part of the photon self energy. Note that in Ref. \cite{mus},
instead of $2/5$ and
$3/5$ as in Eq.(\ref{factor}), these factors are found to be $3/7$ and $4/7$ respectively.
Apart from this minor discrepancy, the use of equilibrium value for $\cal R$
is incorrect for the following reason even though Eq.(\ref{factor}) is approximately
correct as mentioned before.
The  factor $\cal R$ contains the square of the matrix elements involving the
product of the terms $(R^2-M_\infty^2)$ and $(P+L)^2-M_\infty^2$ in the denominator.
Due to the presence of two very near by poles, the matrix element will be
enhancement by an factor of $\sim p^2/M_\infty^2$ while the enhancement
would have been logarithmic if only one of the terms appear in the denominator.
Therefore, there is an
additional enhancement $\sim (2~\lambda_g+\lambda_q)/3$
if the
plasma is unsaturated. This additional enhancement will be compensated
to a large extent by the suppression factor given in  Eq.(\ref{factor}) above
particularly when the plasma is gluon dominated.
In addition to taking out the fugacity factors from the distribution
functions appearing in Eq.(\ref{rela1}), the square of the amplitude
should
also be calculated properly for a non-equilibrated plasma.
Although a detailed calculation needs to be carried out,
naively, $\cal R$ should differ from  the equilibrium
value by a factor of $\sim \lambda_g^{-1}$ when $\lambda_q \ll \lambda_g\ll 1$.
A comparison with Eq.(\ref{factor})
suggests that the photon production rate for bremsstrahlung should have strong
dependence on $\lambda_q$
which (within above approximations) is consistent with the results obtained in section
III. Based on the similar arguments, it can be shown that the AWS photon production will  depend
only on $\lambda_q^2$.

\subsection {Suppression and Enhancement mechanism}

Let us now summarize why the suppression and partial compensation occur
when the plasma is
unsaturated. First, it may be easier to understand from the kinetic theory
arguments. For example, consider the case of bremsstrahlung due to quark-quark
scattering [see Eq.(\ref{rela1}) and Eq.(\ref{identity1})]. The suppression is due to the
unsaturated distribution
functions $n_{_F}(r_0+l_0)$ and $n_{_F}(k_0)$ that appear in the initial
states. Similar suppression occurs in case of quark-gluon scattering except
$n_{_F}(k_0)$ has to be
replaced by $n_{_B}(k_0)$. The above suppression
is partially compensated by an additional enhancement factor that arise due to
the mass effect in the matrix element. It may be noted here that
the gluon distribution that appears
in the initial and final states comes through cutting the effective gluon
propagator. In the field theoretical description, $n_{_B}(k_0)$ does not appear
explicitly, but contained in the effective gluon propagator through the thermal
gluon mass $m_g^2$. Further, notice that the soft gluon associated with $n_{_B}(l_0)$
and responsible for scattering has no role in the above suppression.

The same thing happens in the field theoretical description as well. Since
$R+L$ and $R$ correspond to the same quark in the initial and final states of
scattering, we assign same fugacity to both $R+L$ and $R$ lines.
This argument is also consistent with the kinetic theory for the fact that
any of the physical processes do not change the quark contents in the final
states. Therefore, it is reasonable to assume the same chemical potential
both for the initial and final quark lines i.e. to both $R+L$ and $R$ lines.
From the
conservation of potential, it follows that $L$ line should have
zero chemical potential or unit fugacity. Therefore, $n_{_B}(l_0)$ still follows the Bose
distribution. Then, how do we understand the
suppression due to gluon fugacity which gets compensated by the additional
linear enhancement ?
Let us consider Eq.(\ref{jtjl}) again. In case of equilibrated plasma, it is shown \cite{arun1}
that Eq.(\ref{jtjl}) depends only on the ratio $m_g^2/M_\infty^2$.
The integrals $J_T$ and $J_L$ diverge if the thermal quark mass $M_\infty$ is
switched off. Therefore, the above
singularity is regularized by this  thermal quark mass. In case of
non-equilibrated plasma, $M_\infty^2$ decreases giving
enhancement in   $J_T$ and $J_L$.
However, this enhancement is compensated by a
simultaneous decrease in $m_g^2$. To be more explicit, $M_\infty^2$
differs from the equilibrium value $(g^2T^2/6)$ by a factor of
$(2\lambda_g+\lambda_q)/3$. Similarly, $m_g^2$ differs from the
equilibrium value $(4g^2T^2/9)$ by a factor of $(3\lambda_g+\lambda_q)/4$ when $N_f=2$. For
$\lambda_g=\lambda_q=1$, the ratio $m_g^2/M_\infty^2$ is 1.33. Under two
extreme conditions, $\lambda_q \ll\lambda_g$ and $\lambda_g\ll\lambda_q$, the
above ratios are 1.5 and 1.0 respectively. The corresponding changes in the
$J_T$ and $J_L$ values are within $5\%$ to $10\%$ of the equilibrated values and
can be ignored. Therefore, it is fair enough to say that the enhancement due
to the decrease in $M_\infty^2$ gets (nearly) compensated by the corresponding decrease
in $m_g^2$. In other words, for a gluon dominated plasma, the collinear
enhancement and
the suppression due to gluon fugacity cancels out leaving $J_T$ and $J_L$
practically unchanged. The only factor that suppresses the yield is the
$p$ integral in case of bremsstrahlung that involves only the quark and anti-quark
fugacities and the prefactor $n_{_B}(E)$ in case of
annihilation with scattering.

\subsection{One loop results for comparison}

For the completeness and also
for comparison, in the following,
we briefly mention our previous results for effective one loop
level for the case of a non-equilibrated plasma at zero baryon density.
The  physical processes  of  photon production
(annihilation  $q {\bar q} \rightarrow g \gamma$ and Compton processes $q
({\bar q})g \rightarrow q ({\bar q}) \gamma$)
in the lowest order
of perturbative expansion $({\cal O}(\alpha \alpha_s))$ are
obtained by cutting the
loop diagram as shown in Figure 1. Since the self
energy  is  IR  divergent  in the soft momentum limit,
a  cut-off  parameter  $k_c^2$ is introduced
to separate  the soft from the hard momenta of the intermediate quark. The
soft part is obtained from a resummed quark  propagator  and it involves  a
thermal  quark  mass which  acts  as  an  infrared  cutoff.
The  hard  part  is obtained from the relativistic kinetic theory from the
expression
 \begin{eqnarray}
&& 2E \frac{dR^{hard}}{d^3q}=\frac {N}{(2 \pi)^8} \int \frac {d^3p_1}{2E_1}
    \frac {d^3p_2}{2E_2}  \frac {d^3p_3}{2E_3}~n_1(E_1)~n_2(E_2)\nonumber\\
  && \times (1 \pm n_3(E_3)) \delta (p_1^\mu +~  p_2^\mu -~  p_3^\mu -~ q^\mu )
    \sum |{\cal M}|^2
\label{photon1}
    \end{eqnarray}
 by carrying out integration above the cut-off \cite{kap,dutta1,trax1}.
 In the above, $n_{1,2,3}$ are the parton  distribution functions
 with plus sign for annihilation and the minus sign for the two Compton processes.
 The total rate can be obtained by adding soft and hard contributions together.
 The cutoff dependence cancels out in the summation.
 It is also found that the total photon rate can  be obtained  from the hard
 part alone by using the lower limit of integration $k_c^2$ equal to $2m_q^2$
 where the thermal quark mass $m_q^2$ is given in Eq. (\ref{thermal quark}).
Therefore, Eq.(\ref{photon1}) can be used to estimate
the rate of photon production using appropriate
distribution functions and thermal quark mass
 for non-equilibrium plasma.
Although the Juttner functions
for parton distributions can be used, we restrict to the factorized form for
convenience. However, our conclusions are independent of the above choice.
Using factorized distributions and
the identity,
  \begin{eqnarray}
  n_1~n_2(1 \pm n_3)&=& \lambda_1\lambda_2\lambda_3
  n_1^{eq}~n_2^{eq}~(1\pm n_3^{eq})\nonumber\\
  &&+\lambda_1\lambda_2(1 -\lambda_3) n_1^{eq}~n_2^{eq}.
  \end{eqnarray}
the above equation can be broken into two parts \cite{trax,dutta1}.
For the first part, one can use the analytic form that can be
obtained using Boltzmann distribution for $n_1$ and $n_2$
\cite{kap}
\begin{eqnarray}
&&\left(2E \frac{dR}{d^3q}\right)_1 = \frac{5\alpha~\alpha_s}{9~ \pi^2}~ T^2~e^{-E/T}\nonumber\\
&&\times\Biggl [~\underbrace{~\lambda_q^2 \lambda_g \left \{
\frac{2}{3}~ln\left(\frac{4~E T}{k_c^2}\right)-1.43 \right\}}_{annihilation}\nonumber\\
&& +~\underbrace{~2\lambda_q^2 \lambda_g \left\{
 \frac{1}{6} ln\left(\frac{4~E T}{k_c^2}\right)+0.0075 \right\}}_{Compton} \Biggr ]
\label{trax1}
\end{eqnarray}
Following Ref. \cite{trax}, the second part can be written as
\begin{eqnarray}
&&\left(2E \frac{dR}{d^3q}\right)_2 = \frac{10\alpha~\alpha_s}{9~ \pi^4}~ T^2~e^{-E/T}\nonumber\\
&&\times\Biggl [\underbrace{\lambda_q^2 (1 - \lambda_g)\Biggl \{- 2 -2 \beta +
2~ln \left(\frac{4~ET}{k_c^2}\right)\Biggr\}}_{annihilation}\nonumber\\
&&+~\underbrace{\lambda_q \lambda_g (1- \lambda_q) \Biggl \{ 1 -2 \beta +
2~ln \left(\frac{4~ET}{k_c^2}\right)\Biggr\}}_{Compton} \Biggr]
\label{trax2}
\end{eqnarray}
where $\alpha$, $\alpha_s$ are electromagnetic and strong coupling constants
respectively and $\beta=0.577$ is the Euler constant.
In the above, the first term is the contribution from the annihilation
whereas the second and third terms are due to Compton like processes.
The total rate is estimated by adding Eq.(\ref{trax1}) and Eq.(\ref{trax2}).
It may be pointed out here that Baier {\it et al.} \cite{bair2} have also estimated
the rate for a non-equilibrated plasma.
Although the order of magnitude is same, the
above expressions are different from the result given in \cite{bair2} due to
Boltzmann approximations for the initial states. We have also compared the
above results with the exact numerical calculations using Eq.(\ref{photon1})
and find good agreement. Therefore, we prefer to retain the
above form for consistency with our previous work \cite{dutta1}.

\section {Result and discussion}

In the following, we estimate numerically photon production rates for various processes.
The crucial aspect of the calculation is the numerical integration of the
$J_T$ and $J_L$ functions which depend sensitively only on the ratio $m_{\rm g}^2/M_\infty^2$.
\begin{center}
\begin{figure}[!h]
\begin{minipage}{6cm}
\psfig{figure=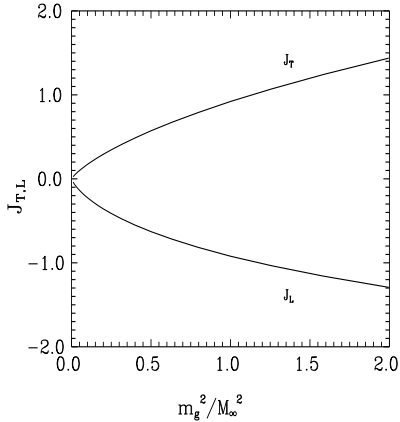,height=6cm,width=6cm}
\end{minipage}
\caption{The variation of $J_T$ and $J_L$ with $m_{\rm g}^2/M_\infty^2$}
\label{fig1}
\end{figure}
\end{center}
Figure \ref{fig1} shows the plot of $J_T$ and $J_L$ over a range of mass
ratios which is of interest in the present study. Although $J_T$ and $J_L$
diverge logarithmically in the limit $M_\infty \rightarrow 0$, it has
approximately a linear dependence in the above region.
In case of equilibrated plasma, the above mass ratio is about $1.33$ (for $N=3$ and $N_f=2$)
and the corresponding $J_T$ and $J_L$ values are found to be
1.108 and -1.064 respectively \cite{priv}.
Note that these values are less exactly by a factor of $4$ than the values
originally reported in \cite{arun1,arun2} and used subsequently by many others.
Since the variation of $m_{\rm g}^2/M_\infty^2$ with fugacity is not significant, we
also use the above values for $J_T$ and $J_L$  both for equilibrated and
non-equilibrated plasma.

Figure \ref{fig2}(a) shows the  comparison between
one and two loop contributions to photon self energy evaluated with
the corrected values of $J_T$ and $J_L$. As in \cite{arun2}
bremsstrahlung dominates in the low momentum region whereas AWS
dominates in the higher momentum scale. Figure \ref{fig2}b shows the
photon production rate at a fixed temperature $T=0.57$ GeV for a chemically
equilibrated plasma ($\lambda_g=\lambda_q=1.0$). The two loop contribution
(bremsstrahlung + AWS) competes or even dominates over one loop
photon production over a wide energy range. Further, it is noticed that
the bremsstrahlung process has strong contribution to photon production
below $E \sim 1 $GeV and falls at a faster rate as compared to the one loop
contribution particularly at higher energy.
Since the
$J_T$ and $J_L$ factors are same, the bremsstrahlung and AWS photon
productions differ only due to different phase space factors. The $p$-integral
in Eq.(\ref{ratejtjl}) involves the quark distribution functions $n_{_F}(p)-n_{_F}(p+E)$
whereas the $p$-integral in Eq.(\ref{ratejtj2}) is nearly independent of
distribution functions for $0<p<E$ and has insignificant contribution for
$p>E$. Therefore, the phase space suppression is stronger for bremsstrahlung
as compared to the AWS process.
The most significant contribution to photon production comes from  the AWS
process at higher photon energies. Our results obtained with the
correct $J_T$ and $J_L$ values are qualitatively in agreement with the
conclusions drawn from the earlier studies of Aurenche {\it et al.}

Next, we consider a chemically unsaturated plasma with two different
initial conditions at RHIC bombarding energy.
Figure \ref{fig3}(a) corresponds to the initial conditions
$T=0.57$ GeV, $\lambda_g=0.09$ and $\lambda_q=0.02$ as obtained from
HIJING model calculation \cite{biro} whereas
Figure \ref{fig3}(b) is plotted with the initial conditions
$T=0.67$ GeV, $\lambda_g=0.34$ and $\lambda_q=0.064$
corresponding  to a typical SSPC model \cite{mus}.
Note that in both cases the initial plasma is gluon dominated.
A general observation is that the AWS contribution is less than the one
loop contributions and the bremsstrahlung seems to be the dominant mechanism
of photon production over a wide range of energy particularly when the
plasma is strongly unsaturated.
The above results can be understood as follows.
The dominant contributions to photon production at the one loop level
comes from the term which is linear in $\lambda_q \lambda_g$ [see
Eq.(\ref{trax1}) and Eq.(\ref{trax2})]. Also see Eq.(45) of Ref. \cite{bair2}.
Since the $J_T$ and $J_L$ are insensitive to fugacities, the contributions
at the two loop level (say) bremsstrahlung is suppressed by a factor of
$\lambda_q$.
\begin{figure}[!h]
\begin{center}
\begin{minipage}{6cm}
\psfig{figure=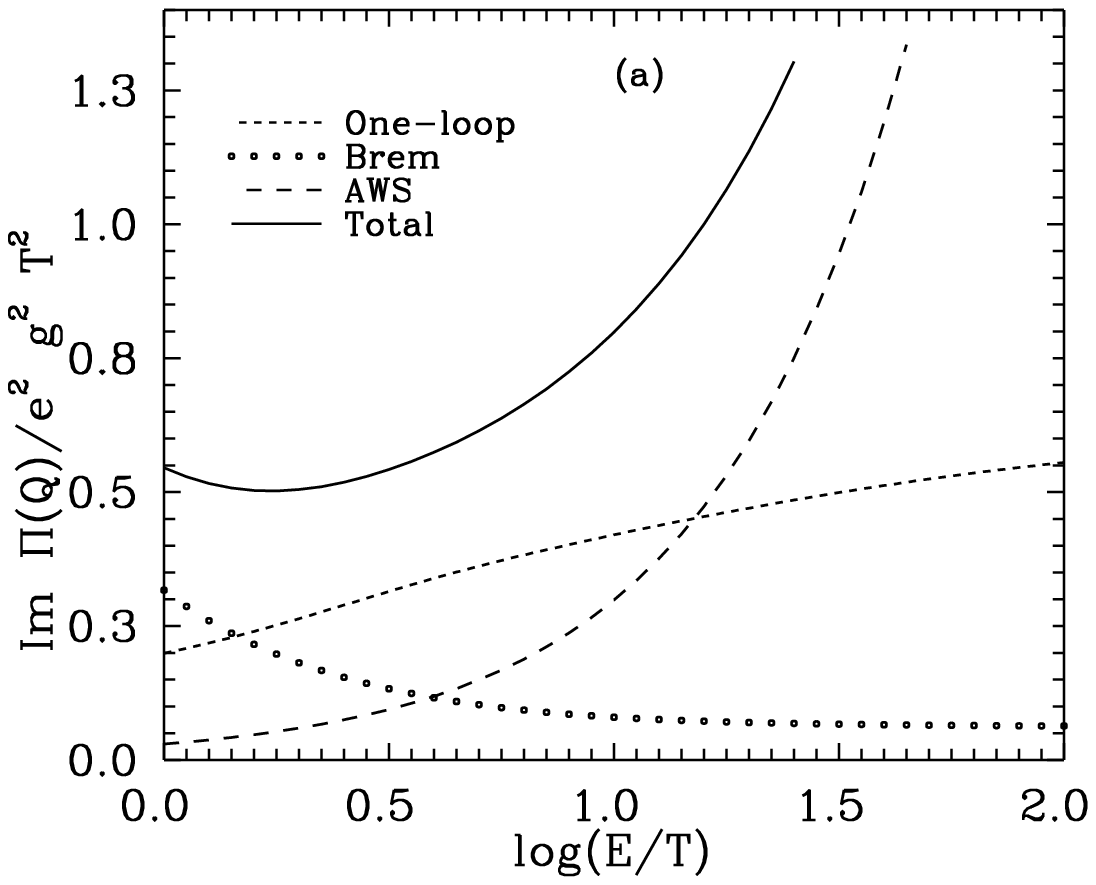,height=6cm,width=6cm}
\end{minipage}
\hspace{1cm}
\begin{minipage}{6cm}
\psfig{figure=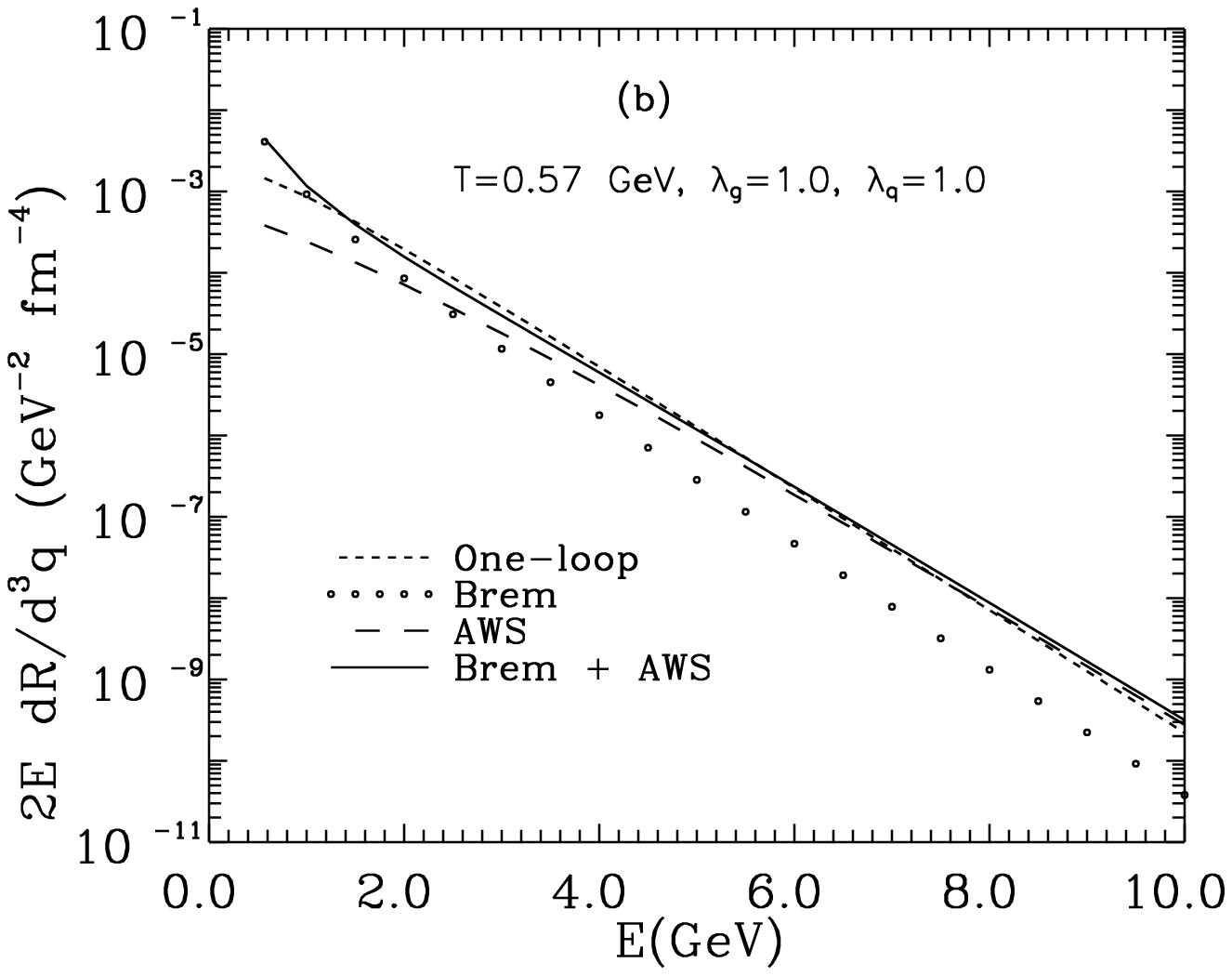,height=6cm,width=6cm}
\end{minipage}
\end{center}
\caption{(a) The photon self energy for one loop and two loop processes as a
function of photon energy.  (b) The rate of photon production for an
equilibrated plasma ($\lambda_g=\lambda_q=1.0$) at constant temperature
$T=0.57$ GeV}
\label{fig2}
\end{figure}
Similarly, the AWS process is suppressed by a factor of $\lambda_q^2$
that arises due to the prefactor although $p$-integral is nearly independent
of any distribution functions. As a consequence, the AWS process is
suppressed strongly as compared to the both one loop and bremsstrahlung
processes. Interestingly, it is the bremsstrahlung which dominates the
photon production at all energies. This is to be contrasted with the
equilibrium situation where the AWS is the dominant mechanism of photon
production at higher energies. Naively, from the kinetic theory
arguments, it is expected that for an unsaturated plasma (gluon dominated), the
bremsstrahlung and the AWS processes will be reduced by a factor of
$\lambda_q\lambda_g$ and $\lambda_q^2\lambda_g$ respectively. However,
the suppression due to $\lambda_g$ gets compensated by the collinear
enhancement which goes up by a factor of $\lambda_g^{-1}$.
These findings are also quite intuitive with processes initiated by one
quark in the initial state being less suppressed than those initiated
by two quarks. Therefore the suppression factors for the AWS, one-loop and
the bremsstrahlung processes are $\lambda_q^2$, $\lambda_q\lambda_g$ and
$\lambda_q$ respectively.
\begin{figure}[ht]
\begin{center}
\begin{minipage}{6cm}
\psfig{figure=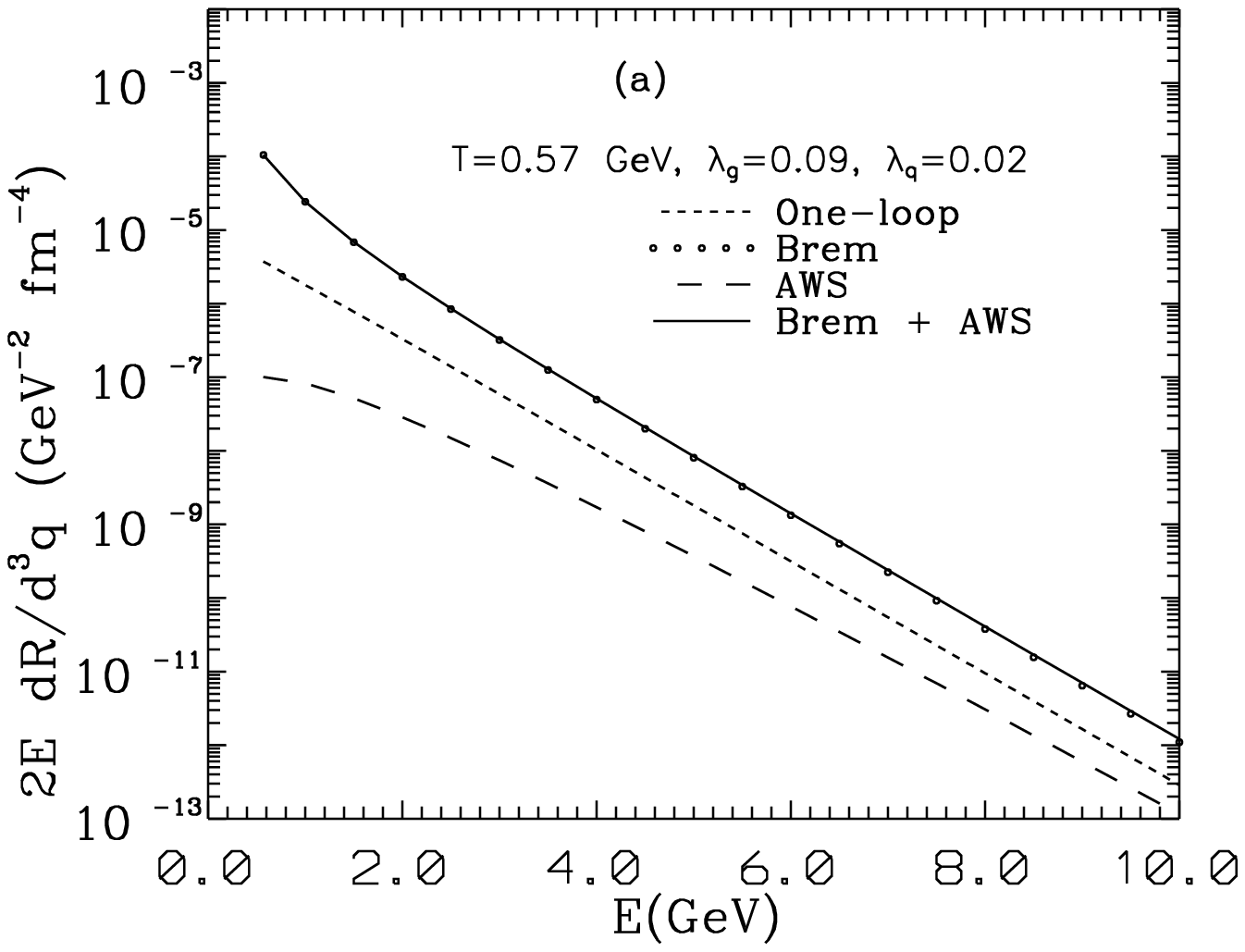,height=6cm,width=6cm}
\end{minipage}
\hspace{1cm}
\begin{minipage}{6cm}
\psfig{figure=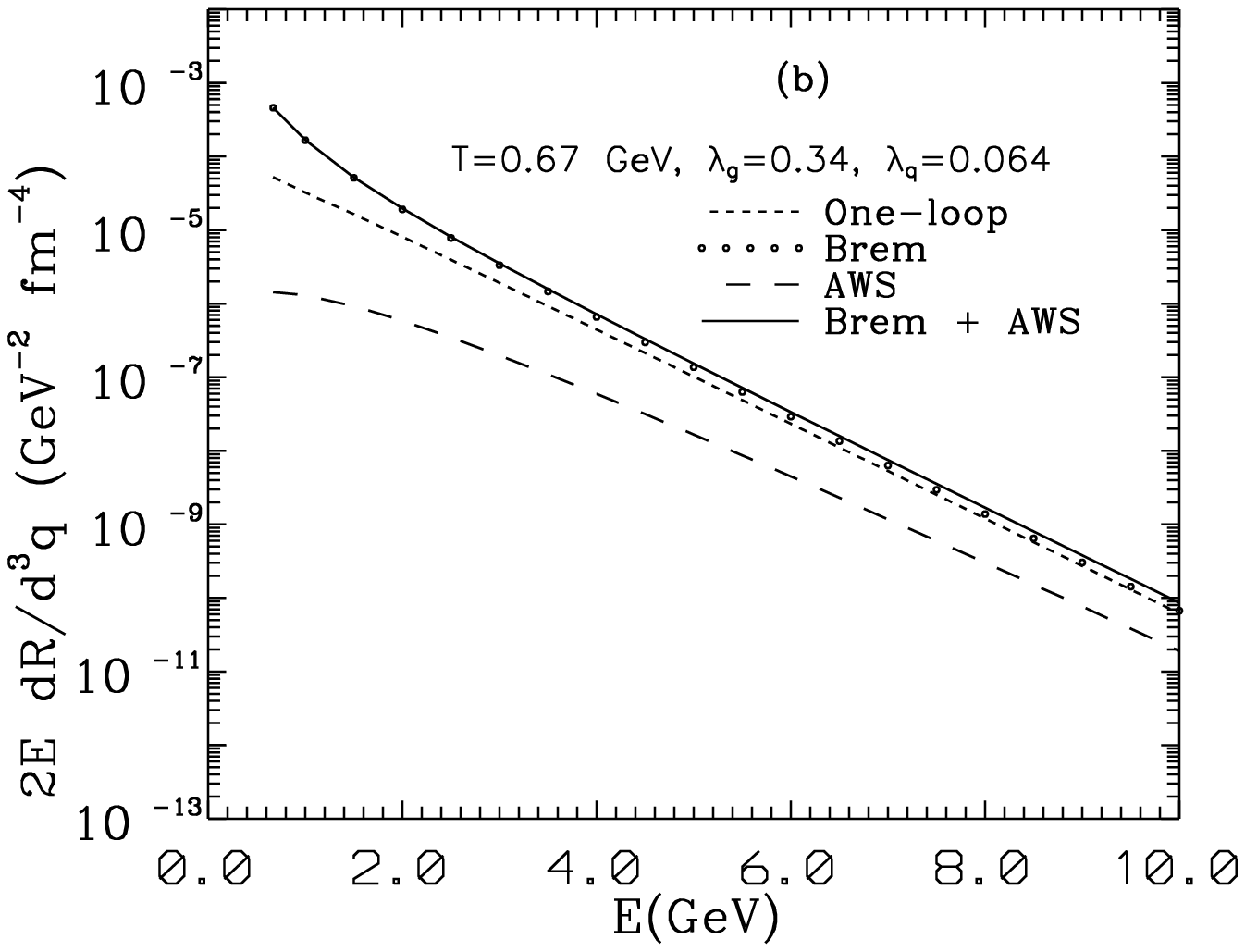,height=6cm,width=6cm}
\end{minipage}
\end{center}
\caption{The rate of photon production at
(a) $T=0.57$ GeV, $\lambda_g=0.09$, $\lambda_q=0.02$
corresponds to HIJING model and
at (b) $T=0.67$ GeV, $\lambda_g=0.34$, $\lambda_q=0.064$
corresponds to SSPC model}
\label{fig3}
\end{figure}

\section {Conclusion}
The effect of chemical potential on photon production from a quark gluon
plasma has been studied. Since, the non-vanishing chemical potential characterizes
an unsaturated phase space, the contributions to photon productions both at
effective one and two loop levels are suppressed as compared to their equilibrium
counterparts. The contributions at the effective two loop level i.e.
the bremsstrahlung and annihilation with scattering (AWS) processes are
suppressed by a factor of $\lambda_q$ and $\lambda_q^2$ respectively. Interestingly,
the above suppressions are found to be independent of the gluon fugacity $\lambda_g$.
The reduction in the photon production rate due to unsaturated gluon distribution
gets compensated to the large extent by the collinear enhancement.
This aspect can be understood
from  the kinetic theory formalism.
However, there is no such enhancement for annihilation and
Compton processes at the one loop level and the rate of photon production
is suppressed by a factor
of $\sim \lambda_q\lambda_g$. Therefore, in case of an unsaturated plasma,
the AWS process is more suppressed  as compared to the one loop
contributions which itself is less by a factor of $\lambda_g$  compared
to the bremsstrahlung process. This is in contrast to the equilibrium
scenario where AWS dominates the photon production followed by one loop and
bremsstrahlung contributions. In either case, whether the plasma is saturated
or unsaturated, the two loop contributions seem to dominate over the one loop
processes particularly at higher photon energies.

Further, we would like to mention
here that an inherent assumption which has gone into the above formalism is
the infinite lifetime of the plasma. As a consequence, the photon production
rate is independent of time and depends only on the photon energy and the temperature
of the plasma. The consideration of the finite life time of the plasma will
lead to the time dependent production rate which may enhance
the photon production further. The finite life time effect has been studied
in \cite{boy} where the plasma is assumed to be both in thermal and chemical
equilibrium. Although the emission rate is a non-equilibrium phenomena, the
present study is quite different in the sense that the non-equilibrium here
refers to a chemically unsaturated plasma that evolves with time. The basic
production rate is still static, but the time dependence arises
due to the hydrodynamical evolution of the plasma.
Therefore, a  meaningful quantity that can be
compared with the experimental results is the space time integrated photon
yields from a plasma undergoing chemical equilibrium. Such a study is being
carried out and will be published elsewhere.

\acknowledgements{}
It is a great pleasure to thank P. Aurenche and F. Gelis for many
fruitful discussions during the course of this work.
We also acknowledge B. Sinha and D. K. Srivastava for
many stimulating discussions.

\appendix
\setcounter{figure}{0}
\renewcommand{\thefigure}{A\arabic{figure}}
\section *{RA formalism at finite chemical potential}

We generalize the RA formalism \cite{arun5,arun6} appropriate
for an unsaturated quark gluon
plasma (QGP) at finite baryon density. Since the QGP is in a thermalized
state, the parton distributions can be described by the Juttner functions
with non-vanishing chemical potential $\mu$. This $\mu$ can be decomposed
as a sum of two components $\mu^c$ and $\mu^b$ where $\mu^c$ characterizes
the unsaturated properties and $\mu^b$ is associated with the finite
baryon density of the plasma.

The propagators in RA formalism are $2\times 2$ diagonal matrices,
constructed  from
the  retarded  and  advanced  propagators  of the T=0 theory while all the
temperature dependence appears in the vertices.
The propagators for fermions and (gauge) bosons, defined  on  the
contour  characterized  by $\sigma$, can be written as,
\begin{eqnarray}
\label{rtf1}
{\cal S}_F(P)&=&(P\sls+M)~~
U^{[\eta]}(P)~D(P)~V^{[\eta]}(P) \\
{G}_{B}^{\mu\nu}(P)&=&-g^{\mu\nu}~~
U^{[\eta]}(P)~D(P)~V^{[\eta]}(P)
\end{eqnarray}
where $[\eta]=B(F)$ for bosonic (fermionic)
propagators and $D(P)$ is the diagonal matrix
\begin{equation}
\label{rtf2}
D(P)=\left(\begin{array}{cc}
\Delta_R(P) & 0\\
0           & \Delta_A(P)
\end{array}\right )
\end{equation}
with the retarded and advanced propagators given by
\begin{equation}
\label{rtf3}
\Delta_{R,A}(P)=\frac{i}{P^2-M^2\pm i\epsilon p_0}
\end{equation}
The matrices $U$ and $V$ are defined as
\begin{equation}
\label{rtf4}
U^{[\eta]}(P)=e^{\beta(p_0-\mu)}n^{[\eta]}(p_0)\left(\begin{array}{cc}
b^{-1} & \eta c^{-1} e^{(\sigma p_0 - x)}\\
b^{-1}e^{-\sigma p_0}           & c^{-1}
\end{array}\right )
\end{equation}
\begin{equation}
\label{rtf5}
V^{[\eta]}(P)=\left(\begin{array}{cc}
b & \eta b e^{(\sigma p_0 - x)}\\
-c e^{-\sigma p_0}           &- c
\end{array}\right )
\end{equation}
where $b$ and $c$ are arbitrary scalar functions of $P$,
$\eta=\pm1$  for a boson(fermion) and
n$^{[\eta]}(p_0)$ is the  usual Bose-Einstein or
Fermi-Dirac distribution defined as
\begin{equation}
\label{distr}
n^{[\eta]}(p_0)=\frac{1}{e^x-\eta};~~~~~x=\beta(p_0-\mu)\nonumber
\end{equation}
$\beta$ being the inverse of temperature and $\mu$ is the
chemical potential as defined above.
In the RA formalism, $U$ is associated to an outgoing line while $V$ is
associated to an incoming line. All the temperature dependence which is
contained in $U$ and $V$ will then appear in the vertices.

The different types of vertices are calculated depending
on the momentum flow. Let us consider a vertex with all lines incoming as
shown in  Figure \ref{vertex1}.
\begin{figure}[ht]
\centerline{\psfig{figure=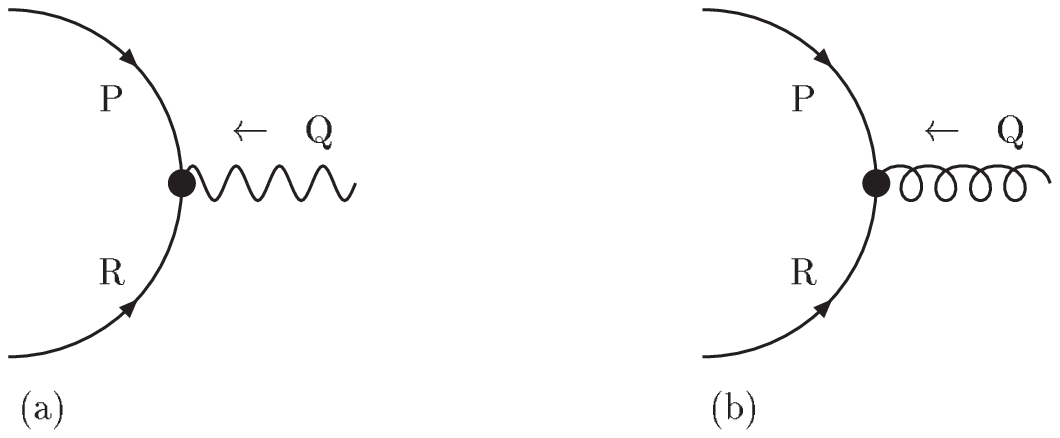,height=3.0cm,width=6cm}}
\caption{(a) EM and (b)Strong vertices}
\label{vertex1}
\end{figure}
The new vertex function has the form
\begin{equation}
-i\gamma_{\abr}(P,Q,R)=-ig_{abd}V^F_{\alpha a}(P)~V^B_{\beta b}(Q)~
V^F_{\rho d}(R)
\label{rtf6}
\end{equation}
The Greek indices take the value R or A and the Latin indices
refer to the $1$ (particle) $2$ (ghost) of the usual formulation of
the real time formalism (RTF) so that $g_{111}=\zeta$, $g_{222}=-\zeta$, where $\zeta$=$e$
or $g$ depending on the electromagnetic
or strong vertex and all other couplings being zero. From the definitions
above it can be shown that,
\begin{eqnarray}
&&\gamma_{\abr}(P,Q,R) = \zeta[b(P)]^{\delta_{\alpha R}}
[b(Q)]^{\delta_{\beta R}}[b(R)]^{\delta_{\rho R}}\nonumber \\
&&\times [-c(P)]^{\delta_{\alpha A}} [-c(Q)]^{\delta_{\beta A}}
[-c(R)]^{\delta_{\rho A}}~e^{\sigma L_0}\nonumber\\
&&\left[1-(-1)^{\delta_{\alpha R}+
\delta_{\rho R}}~e^{-\beta~L_0^\prime}\right]
\label{rtf7}
\end{eqnarray}
with $L_0=p_0\delta_{\alpha R} +q_0 \delta_{\beta R} +
r_0 \delta_{\rho R}$
and  $L_0^\prime=[p_0-\epsilon(P)\mu_P]\delta_{\alpha R} +[q_0-\epsilon(Q)
\mu_Q] \delta_{\beta R} +
[r_0-\epsilon(R)\mu_R] \delta_{\rho R}$.
In the above, with each momentum $P$, $Q$ and $R$, we have introduced
an associated chemical
potential $\epsilon(P)\mu_P$, $\epsilon(Q)\mu_Q$ and $\epsilon(R)\mu_R$. The
sign function has been introduced to ensure that when
the momentum reverses, the
associated chemical potential also changes its sign. This aspect is also consistent
with the definition of parton distribution functions as given in section II.
The causality requirement that three particles propagating forward in time
(or backward in time) can not annihilate into (or be created from) the vacuum
demands that $\gamma_{AAA}$ and $\gamma_{RRR}$ should vanish. It is immediately
clear from Eq.(\ref{rtf7}) that $\gamma_{AAA}$ always vanishes. However,
the vanishing of $\gamma_{RRR}$ requires  energy and chemical potential conservation.
In case of finite baryon density, both $\mu^c$ and $\mu^b$ needs
to be conserved separately.
Therefore, the following set of conservation equations
are satisfied when $\gamma_{RRR}=0$
\begin{eqnarray}
&&p_0+q_0+r_0=0;\nonumber\\
&&\mu_P^c+\mu_Q^c+\mu_R^c=0;\nonumber\\
&&\mu_P^b+\mu_R^b=0
\label{rtf8}
\end{eqnarray}
Note that in the above the baryo-chemical potential for photon or gluon has been
set to zero.
Next, we consider a crossing fermion line as shown in Figures \ref{vertex2}(a)
and \ref{vertex2}(b)
with the conservation laws,
\begin{eqnarray}
P+Q=R;~~
\mu_P^c+\mu_Q^c=\mu_R^c;~~
\mu_P^b=\mu_R^b
\label{rtf9}
\end{eqnarray}
The vertex  function can be evaluated from
\begin{equation}
-i\gamma_{\abrp}(P,Q;R)=-ig_{abd}V^F_{\alpha a}(P)~V^B_{\beta b}(Q)~
U^F_{d \rho}(R)
\label{rtf10}
\end{equation}
Using the definition of $U$ and $V$, the above equation can be
written similar way as that of Eq.(\ref{rtf7}) given by,
\begin{eqnarray}
&&\gamma_{\abrp}(P,Q;R) = \zeta[b(P)]^{\delta_{\alpha R}}
[b(Q)]^{\delta_{\beta R}}[b(R)]^{-\delta_{\rho R}}\nonumber \\
&&\times [-c(P)e^{-\sigma p_0}]^{\delta_{\alpha A}}
[-c(Q)e^{-\sigma q_0}]^{\delta_{\beta A}}
[-c(R)e^{-\sigma r_0}]^{-\delta_{\rho A}}\nonumber\\
&&~n_{_F}(r_0)[e^{\beta(r_0-\mu_R)}]^{\delta_{\rho R}}
\left[(-1)^{\delta_{\rho R}}+(-1)^{\delta_{\alpha R}}
~e^{-\beta P_0^\prime}\right]
\label{rtf11}
\end{eqnarray}
where,  $P_0^\prime=
[p_0-\epsilon(P)\mu_P]\delta_{\alpha R} +[q_0-\epsilon(Q)\mu_Q] \delta_{\beta R} -
[r_0-\epsilon(R)\mu_R] \delta_{\rho A}$.
\begin{figure}[ht]
\centerline{\psfig{figure=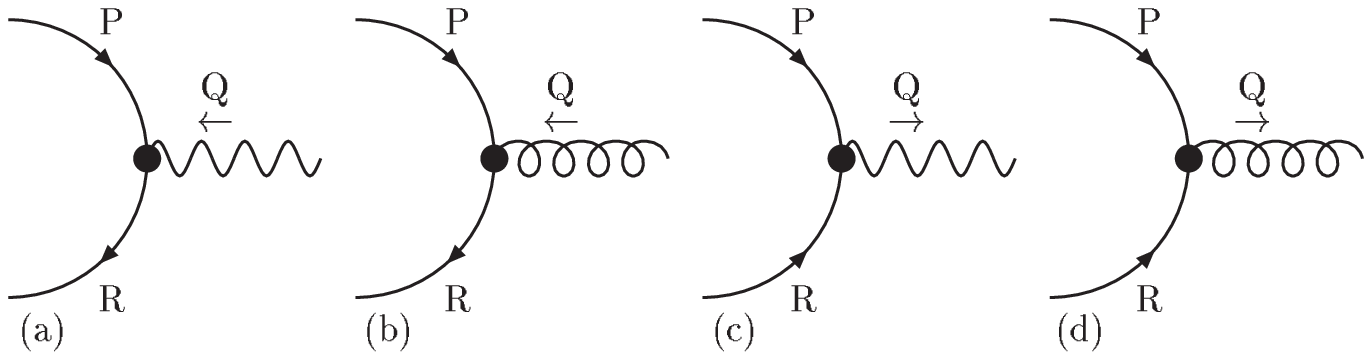,height=3cm,width=6.5cm}}
\caption{
(a) and (b) are the  QED and QCD vertices for crossing
fermion line
(c) and (d) are the  QED and QCD vertices for crossing
boson line}
\label{vertex2}
\end{figure}
By comparing with Eq.(\ref{rtf7}), we can derive the relations
\begin{equation}
\gamma_{\abRp}(P,Q;R) =-\frac{e^{-\sigma r_0}~n_{_F}(-r_0)}
{b(R)c(-R)}\gamma_{\abA}(P,Q,-R)
\label{rtf12}
\end{equation}
\begin{equation}
\gamma_{\abAp}(P,Q;R) =-\frac{e^{\sigma r_0}~n_{_F}(r_0)}
{b(-R)c(R)}\gamma_{\abR}(P,Q,-R)
\label{rtf13}
\end{equation}
The choice
\begin{equation}
 b(-R)c(R)=-n_{_F}(r_0)e^{\sigma r_0}
\label{rtf14}
\end{equation}
gives the crossing relation for fermion
\begin{equation}
\gamma_{\abrp}(P,Q;R) = \gamma_{\abrb}(P,Q,-R)
\label{rtf15}
\end{equation}
where ${\bar \rho}=A,R$ is the conjugate index of $\rho=R,A$.
The crossing property of boson [see Figures \ref{vertex2}(c)
and \ref{vertex2}(d)] can also be derived in a similar way
except the conservation
\equation P+R=Q;~~
\mu_P^c+\mu_R^c=\mu_Q^c;~~
\mu_P^b+\mu_R^b=0
\endequation
should be followed. The replacement of $-n_F(r_0)$ in Eq.(\ref{rtf14})
with $n_{_B}(q_0)$ leads to the equation for boson
\begin{equation}
 b(-Q)c(Q)=n_{_B}(q_0)e^{\sigma q_0}.
\label{rtf16}
\end{equation}
It needs to be stressed here that the factor $n_{_B}(q_0)$ represents the
boson distribution function of energy $q_0$ and with appropriate chemical
potential $\mu_Q$. This distribution is  not to be associated with the
emitted photon of energy $q_0$ which has no chemical potential and
distribution function.
Using Eq.(\ref{rtf7}) with the conditions given by Eq.(\ref{rtf14}) and
Eq.(\ref{rtf16}) and also with the choice $b=1$, all the required vertices
can be calculated. For example, we consider Figure \ref{awsfig} which contributes
to the physical process annihilation with scattering.
\begin{figure}[ht]
\centerline{\psfig{figure=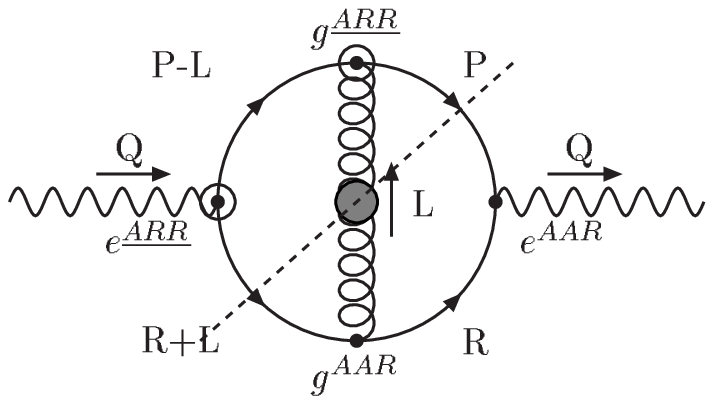,height=3cm,width=6cm}}
\caption{Self energy which contribute to AWS process}
\label{awsfig}
\end{figure}
The Eq.(\ref{rtf7}) can be simplified for $e^{AAR}$
using appropriate conservation laws to get
\begin{equation}
e^{ AAR}(R,P,-Q)=-\frac{e~n_{_F}(r_0)n_{_F}(p_0)}{n_{_B}(q_0)}
\label{rtf17}
\end{equation}
which has same form what one would have expected for an equilibrated case
except the distribution functions which now contain appropriate chemical
potential. Assuming $\mu_P^c=\mu_R^c=\mu_q^c$ (quark chemical potential)
and $|\mu_P^b|=|\mu_R^b|=\mu_q^b$,
the conservation laws, $Q=P+R;~~\mu_Q^c=\mu_P^c+
\mu_R^c$ and $\mu_P^b+\mu_R^b=0$ suggest that the boson line should have
a total chemical potential $2\mu_q^c$ while the total chemical potential for
$P$ and $R$ lines are $\mu_q^c+\mu_q^b$ and $\mu_q^c-\mu_q^b$.
(Note that in this topology, the chemical potential $\mu_R$ and $\mu_P$ are
associated with a quark and anti-quark respectively. Both should have same
chemical potential $\mu^c$ and opposite baryo-chemical potential $\mu^b$).
Using the distribution
function with the above chemical potentials, Eq.(\ref{rtf17}) can also be
written as
\begin{equation}
e^{ AAR}(R,P,-Q)=e[n_{_F}(r_0)-n_{_F}(-p_0)]
\label{rtf18}
\end{equation}
Similarly, we can write other vertices
\begin{eqnarray}
&&g^{ AAR}(R+L,-L,-R)=g[n_{_B}(l_0)+n_{_F}(r_0+l_0)]\nonumber\\
&&e^{\underline{ ARR}}(Q,-P+L,-R-L)\nonumber\\
&&=-e^{ ARR}(Q,-P+L,-R-L)=-e\nonumber\\
&&g^{\underline{ ARR}}(P-L,-P,L)=-g^{ ARR}(P-L,-P,L)=-g\nonumber\\
\label{rtf19}
\end{eqnarray}
Since the chemical potential associated with $R+L$ and $R$ are same
$(\mu=\mu_q^c+\mu_q^b)$, the chemical potential for $L$ line is zero.
The above diagram contributes to bremsstrahlung when $P$ reverses it's direction.
In this topology, the total chemical potential for $R$, $P$ and $Q$ lines are
$\mu_q^c+\mu_q^b$, $\mu_q^c+\mu_q^b$ and $0$ respectively. The corresponding four vertices are
\begin{eqnarray}
&&e^{ AAR}(R,-P,-Q)=e[n_{_F}(r_0)-n_{_F}(p_0)]\nonumber\\
&&g^{ AAR}(R+L,-L,-R)=g[n_{_B}(l_0)+n_{_F}(r_0+l_0)]\nonumber\\
&&e^{\underline{ ARR}}(Q,P+L,-R-L)=-e^{ ARR}(Q,P+L,-R-L)\nonumber\\
&&=-e\nonumber\\
&&g^{\underline{ ARR}}(-P-L,P,L)=-g^{ ARR}(-P-L,P,L)=-g\nonumber\\
\label{rtf20}
\end{eqnarray}

\end{document}